\documentclass[preprint,prd,aps,nofootinbib,tightenlines,superscriptaddress]{revtex4}

\usepackage{graphicx}
\usepackage{amsmath}
\usepackage{amssymb}
\usepackage{bm}
\usepackage{graphics}
\usepackage{xspace}

  \newcount\hour \newcount\minute
  \hour=\time \divide \hour by 60
  \minute=\time
  \newcommand{\mydate}{\ \today \ - \number\hour :\ifnum \minute<10 0\fi 
\number\minute}

\def\OMIT#1{}

\newcommand{\nn}{\nonumber} 
\newcommand{\eq}[1]{Eq.~\eqref{eq:#1}}
\newcommand{\eqs}[2]{Eqs.~\eqref{eq:#1} and \eqref{eq:#2}}
\newcommand{\bn}{{\bar n}}

\newcommand{\lqcd}{\Lambda_\mathrm{QCD}}
\newcommand{\plus}{\ensuremath{\! + \!}}
\newcommand{\minus}{\ensuremath{\! - \!}}
\newcommand{\ECM}{E_\mathrm{cm}}

\newcommand{\beq}{\begin{equation}}
\newcommand{\eeq}{\end{equation}}
\newcommand{\beqa}{\begin{eqnarray}}
\newcommand{\eeqa}{\end{eqnarray}}

\newcommand{\nslash}{n\!\!\!\slash}
\newcommand{\bnslash}{\bar{n}\!\!\!\slash}
\newcommand{\df}{{\mathrm{d}}}

\newcommand{\fjet}{fragmenting jet function\xspace}
\newcommand{\SCETa}{\ensuremath{{\rm SCET}_{\rm I}}\xspace}
\newcommand{\SCETb}{\ensuremath{{\rm SCET}_{\rm II}}\xspace}

\begin{document}
\setlength\baselineskip{17pt}


\preprint{ \vbox{ \hbox{MIT-CTP 4063}
\hbox{TUM-EFT 3/09} 
 \hbox{arXiv:0911.4980}
 \hbox{\today}
} 
}

\title{\boldmath \Large
Quark Fragmentation within an Identified Jet
} 

\vspace*{1cm}

\author{Massimiliano Procura}
\affiliation{Center for Theoretical Physics, Laboratory for Nuclear Science,
Massachusetts Institute of Technology, Cambridge, MA 02139}
\affiliation{Physik-Department, Technische Universit\"at M\"unchen, D-85748
  Garching, Germany \vspace{1cm}}
\author{Iain W.~Stewart}
\affiliation{Center for Theoretical Physics, Laboratory for Nuclear Science,
Massachusetts Institute of Technology, Cambridge, MA 02139}



\begin{abstract}
\vspace*{0.3cm}
  
We derive a factorization theorem that describes an energetic hadron $h$
fragmenting from a jet produced by a parton $i$, where the jet invariant mass is
measured. The analysis yields a ``\fjet'' ${\cal G}_i^h(s,z)$ that depends on
the jet invariant mass $s$, and on the energy fraction $z$ of the fragmentation
hadron. We show that ${\cal G}^h_i$ can be computed in terms of perturbatively
calculable coefficients, ${\cal J}_{ij}(s,z/x)$, integrated against standard
non-perturbative fragmentation functions, $D_j^{h}(x)$. We also show that
$\sum_h \int dz\,z\, {\cal G}_i^h(s,z)$ is given by the standard inclusive jet
function $J_i(s)$ which is perturbatively calculable in QCD. We use
Soft-Collinear Effective Theory and for simplicity carry out our derivation for
a process with a single jet, $\bar{B} \to X h \ell \bar{\nu}$, with invariant
mass $m_{X h}^2 \gg \Lambda_{\rm QCD}^2$. Our analysis yields a simple
replacement rule that allows any factorization theorem depending on an inclusive
jet function $J_i$ to be converted to a semi-inclusive process with a
fragmenting hadron $h$. We apply this rule to derive factorization theorems for
$\bar B\to X K \gamma$ which is the fragmentation to a Kaon in $b\to s\gamma$,
and for $e^+e^- \to \mbox{(dijets)}+h$ with measured hemisphere dijet invariant
masses.
 
\end{abstract}

\maketitle


\newpage

\section{Introduction} \label{sec:introduction}
Factorization theorems are crucial for applying QCD to hard scattering processes
involving energetic hadrons or identified jets. In single inclusive hadron production, an  initial
energetic parton $i=\{u,d,g,\bar u,\ldots\}$ produces an energetic hadron $h$ and
accompanying hadrons $X$.
Factorization theorems for these fragmentation processes have been derived at leading power for
high-energy $e^+e^-\to Xh$,
\begin{align} \label{eq:fact1}
  d\sigma = \sum_i  d\hat\sigma_i \otimes D_i^h \,,
\end{align}
as well as lepton-nucleon deeply inelastic scattering, $e^- p \to e^- Xh$,
\begin{align} \label{eq:fact2}
  d\sigma = \sum_{ij} d \hat\sigma_{ij} \otimes D_i^h \otimes f_{j/p} \,.
\end{align}
For a factorization review see Ref.~\cite{Collins:1989gx}.  In
\eqs{fact1}{fact2} the cross sections are convolutions of perturbatively
calculable hard scattering cross sections, $d\hat\sigma$, with non-perturbative
but universal fragmentation functions $D_i^h(z)$, and parton distributions
$f_{j/p}(\xi)$.  The fragmentation functions $D_i^h(z)$ encode information on
how a parton $i$ turns into the observed hadron $h$ with a fraction $z$ of the
initial parton large momentum.  Fragmentation functions are also often used for
processes where a complete proof of factorization is still missing, such as
high-energy hadron-hadron collision, $H_1 H_2 \to h X$.

Another interesting class of hard scattering processes are those with identified
jets. Examples include dijet production $e^+e^- \to X_{J_1} X_{J_2} X_s$ where
$X_{J_{1,2}}$ are two jets of hadrons, and $X_s$ denotes soft radiation between
the jets. If we measure an inclusive event shape variable such as thrust, or
hemisphere invariant masses, then the cross section for this dijet process has
the leading order factorization
theorem~\cite{Korchemsky:1999kt,Korchemsky:2000kp,Fleming:2007qr,Schwartz:2007ib}
\begin{align} \label{eq:fact3}
  d\sigma = H_{\rm 2jet}\:  J\otimes J \otimes S_{\rm 2jet} \,.
\end{align}
Here the $J=J(s)$ are inclusive jet functions depending on a jet invariant mass
variable $s$, $S_{\rm 2jet}$ is a soft function which gets convoluted with the
$J$s as denoted by $\otimes$, and $H_{\rm 2jet}$ is a multiplicative hard
coefficient.  Another example of this type is the $\bar{B}\to X_u
\ell\bar\nu_\ell$ decay in a region of phase space where $X_u$ is jet-like
($\Lambda_{\rm QCD} \ll m_{X_u}^2 \ll m_B^2$). Here the leading order factorization
theorem for the decay rate is~\cite{Korchemsky:1994jb,Bauer:2001yt}
\begin{align} \label{eq:fact4}
  d\Gamma &= H \: J_1 \otimes S \,,
\end{align}
with a hard function $H$ for the underlying $b\to u\ell\bar\nu_\ell$ process,
the same inclusive jet function $J$ as in the previous example, and a ``shape
function'' $S$ which is the parton distribution for a $b$-quark in the $B$-meson
in the heavy quark limit.

In this paper we will analyze processes which combine the above two cases,
namely both the fragmentation of a hard parton $i$ into $h$ and the measurement
of a jet invariant mass. Since this probes fragmentation at a more differential
level, we expect it can teach us interesting things about the jet dynamics
involved in producing $h$, and shed light on the relative roles of perturbative
partonic short-distance effects and non-perturbative hadronization. We derive
factorization theorems that depend on a new ``\fjet" ${\cal G}_i^h(s,z)$. This
function depends on $s$, the jet invariant mass variable, and on $z$, the ratio
of the large light-like momenta of the fragmentation hadron and parton.  Two
interesting formulae involving ${\cal G}_i^h$ will be derived. The first formula
states that 
\begin{align} \label{eq:JisG}
  J_i(s,\mu) &= \frac12\: \sum_h \int  \frac{dz}{(2\pi)^3}  
   \: z \: {\cal G}_i^h(s,z,\mu) \,,
\end{align}
so that the inclusive jet-function can be decomposed into a sum of terms, ${\cal
  G}_i^h$, for fragmentation to a hadron $h$ with $m_h^2 \ll m_X^2$. This
formula also leads to a replacement rule for factorization theorems, where we
can take any process involving an inclusive jet function, and replace $J_i\to
{\cal G}_i^h$ to obtain the corresponding process with a fragmenting jet.

The second formula states that to leading order in $\lqcd^2/s \ll 1$ we have
\begin{align} \label{eq:GisD}
 {\cal G}_i^h(s,z,\mu) &= \sum_j \int \frac{dx}{x} \,
   {\cal J}_{ij}\Big(s,\frac{z}{x},\mu\Big)\,
   D_{j}^h(x,\mu) \,,
\end{align}
so that the \fjet can be expressed in terms of perturbatively calculable
coefficients ${\cal J}_{ij}$, together with the standard unpolarized fragmentation
functions $D_{j}^h(x,\mu)$ renormalized in the $\overline{\rm MS}$ scheme.

To introduce the concept of ${\cal G}_i^h$ and study its properties, we will
specialize to a process with a single jet recoiling against leptons, namely
$\bar B\to X h \ell\bar\nu_\ell$.  Using Soft-Collinear Effective Theory (SCET)
\cite{Bauer:2000ew,Bauer:2000yr,Bauer:2001ct,Bauer:2001yt} we derive
leading-order factorization formulae for $\bar{B} \to X h \ell \bar{\nu}_\ell$
decay rates, in the region of phase space characterized by $\Lambda_{\rm QCD}^2
\ll m_{X h}^2 \ll m_B^2$ where the hadronic final state is jet-like, and where
the energetic hadron $h$ fragments from the jet. This $b\to u\ell\bar{\nu}_\ell$
process has the virtue of having a single jet whose invariant mass can be
measured in a straightforward manner with available $B$-factory data. Despite
our focus on $\bar B\to Xh\ell\bar \nu_\ell$ the results obtained can be
immediately generalized to fragmentation in other processes where a jet
invariant mass measurement is made. Two examples will be described.

The paper is organized as follows. In Sec.~\ref{sec:dfunc} we review the
standard definition of the quark fragmentation function $D_q^h(z)$ and highlight
features that are relevant for later parts of our analysis.
Sec.~\ref{sec:BXhlnu} is devoted to the process $\bar B\to X h
\ell\bar\nu_\ell$, including a discussion of kinematics in Sec.~\ref{sec:kin}.
Results for relevant differential decay rates in terms of components of an
appropriate hadronic tensor are given in Sec.~\ref{sec:decay}.
Sec.~\ref{sec:scetsec} contains the derivation of the SCET factorization
formulae for $\bar B\to Xh \ell\bar\nu_\ell$, and the definition of the ``\fjet"
${\cal G}_i^h$.  In Sec.~\ref{sec:relations} we discuss the relations shown
above in \eqs{JisG}{GisD}.  Conclusions, outlook, and the generalization to
other processes are given in Sec.~\ref{sec:conclusions}.

\section{The fragmentation function $D(z)$} \label{sec:dfunc}

Defining $n^\mu=(1,0,0,1)$ and $\bn^\mu=(1,0,0,-1)$, the light-cone components
of a generic four-vector $a^\mu$ are denoted by $a^+=n \cdot a$ and $a^-=
\bar{n} \cdot a$ where $n^2=\bar{n}^2=0$ and $n \cdot \bar{n}=2$. With $a_\perp^\mu$
we indicate the components of $a^\mu$ orthogonal to the plane spanned by $n^\mu$
and $\bar{n}^\mu$. For energetic collinear particles we will follow the
convention where the large momentum is $p^-$ and the small momentum is $p^+$.

Let us consider a quark $q$ with momentum $k^\mu$ fragmenting to an observed
hadron $h$ with momentum $p^\mu$. In a frame where $\vec k_{\perp}=0$, the
hadron has $p_h^- \equiv z\, k^-$ and $p_h^+=(\vec{p}_h^{\,\perp\; 2} +
m_h^2)/p_h^-$.  The standard unpolarized fragmentation function $D_i^h(z)$ is
defined as the integral over $p_h^\perp$ of the ``probability distribution''
that the parton $i$ decays into the hadron $h$ with momentum
$p_h^\mu$~\cite{Collins:1981uw,Collins:1992kk}, see also \cite{Mueller:1978xu, Georgi:1977mg, Ellis:1978ty,
  Curci:1980uw}. With
the gauge choice $\bar{n} \cdot A=0$, the unrenormalized quark fragmentation
function has the following operator definition~\cite{Collins:1981uw}:
\begin{align} \label{eq:defD} 
 D_q^h(z)&=\frac{1}{z}\int\!\! \df^2 p_h^\perp 
\int\! \frac{\df x^+\,\df^2 x_\perp}{ 2(2 \pi)^3}\;e^{\,i k^-  x^+/2} 
\frac{1}{4 N_c}\,{\rm Tr} \sum_X 
  \langle 0 |  \bnslash\, \psi (x^+, 0, x_\perp )| X h
  \rangle \langle X h | \bar{\psi}(0) |0 \rangle \Big|_{p_{X h}^\perp=0}~,
\end{align}
where $\psi$ is the quark field quantized on $x^-=0$, $N_c=3$ is the number of
colors, and the trace is taken over color and Dirac indices. In \eq{defD} the
state $|Xh\rangle = | X h(p_h)\rangle$ has a hadron $h$ with momentum
$p_h=(p_h^-,{\vec p}_h^{\,\perp})$, and an average over polarizations of
$h$ is assumed.  Boost invariance along the non-$\perp$ direction implies
that $D_q^h$ can only be a function of $z=p_h^-/k^-$ and not $k^-$ or $p_h^-$
individually.

Performing a rotation and a boost to a frame where $\vec p_h^{\,\perp} =0$ with
$p_h^-$ left unchanged, $\vec k_{\perp}$ becomes $-\vec p_h^{\,\perp}/z$, and
\eq{defD} can be written in a gauge-invariant form as~\cite{Collins:1981uw}
\beq   \label{eq:defD2} 
D_q^h(z) = z \int \frac{\df x^+}{4\pi}\;e^{\,i k^-
  x^+/2}\, \frac{1}{4 N_c}\,{\rm Tr} \sum_X \left. 
  \,\langle 0| \bnslash\, \Psi (x^+, 0, 0_\perp )| X h \rangle \langle X h
  | \bar{\Psi}(0) |0 \rangle \right|_{p_h^\perp=0} \,,
\eeq
where the field $\Psi(x^+)=\Psi(x^+,0,0_\perp)$ contains an anti path-ordered
Wilson line of gluon fields, in a $\bar 3$ representation
\beq \Psi(x^+) \equiv \psi(x^+) \Big[ \bar {\rm P}\, {\rm exp} \left( i g
  \int_{x^+}^{+\infty} \!\!\df s\: \bar{n} \cdot A^T(s \bar{n}) \right) \Big] \,.
\eeq
We note that the form of \eq{defD2} is not altered if we perform a Lorentz
transformation to a frame where $\vec p_h^{\,\perp}$ equals an arbitrary fixed
reference value $\vec p_\perp^{\;\rm ref}$. In this case $\vec k_\perp = (\vec
p_\perp^{\;\rm ref}-\vec p_h^{\,\perp})/z$ and $\vec p_\perp^{\;\rm ref}$ does not
play any role due to the integrals over $p_h^\perp$ and $x_\perp$ in \eq{defD}.

Our knowledge of the fragmentation functions is anchored to the use of factorization
theorems to describe measurements of single-inclusive high-energy processes.
Constraints are obtained by using perturbative results for the
partonic hard collision as input.  For example, writing out the complete form of
\eq{fact1} for single-inclusive $e^+ e^-$ annihilation into a specific hadron
$h$ at center-of-mass energy $\ECM$, we have
\beq \label{eq:eeXh}
\frac{1}{\sigma_0}\,\frac{d \sigma^h}{d z} \big( e^+ e^- \to h\, X \big)= \sum_i
\int_z^1 \frac{d x}{x}\;C_i(\ECM, x, \mu)\; D_i^h(z/x, \mu) \,,
\eeq
where $\sigma_0$ is the tree level cross section for $e^+e^-\to \text{hadrons}$,
$\mu$ is the renormalization scale in the $\overline{\rm MS}$ scheme, and $C_i$
is the coefficient for the short-distance partonic process producing the
parton $i$.  In \eq{eeXh} the sum includes the contributions from the different
parton types, $i=u,\, \bar{u},\,d,\,g \dots$ and the $C_i$'s are calculable in
perturbation theory, so measurements of $d\sigma^h/dz$ constrain $D_i^h$.

Model parameters for fragmentation functions have been extracted by fitting
to cross section data for single charged hadron inclusive $e^+ e^-$
annihilation, including high statistics measurements at CERN-LEP and
SLAC~\cite{Kretzer:2000yf,Kniehl:2000fe,Albino:2005me,Hirai:2007cx}. More
recently, these data have been combined with semi-inclusive lepton-nucleon
deeply inelastic and $pp$ cross sections from HERMES and RHIC experiments,
respectively, to perform a global analysis of pion, kaon and (anti-)proton
fragmentation based on the factorized expressions for the relevant cross
sections, with partonic input at next-to-leading order in QCD perturbation
theory~\cite{deFlorian:2007aj,deFlorian:2007hc}, see also~\cite{Albino:2008fy}.
These analyses confirm the universal nature of the fragmentation function, and,
for the $\pi^+$, constrain the fragmentation model for the dominant
$D_{u}^\pi(z)$ with uncertainties at the 10\% level for $z \gtrsim
0.5$~\cite{Hirai:2007cx}.  There is less sensitivity to gluon fragmentation
functions, and, correspondingly, these have larger uncertainties.

Factorization theorems like the one in \eq{eeXh} have been proven to all orders
in $\alpha_s$ at leading order in $\lqcd/\ECM$ for processes in which all
Lorentz invariants like $\ECM^2 = (p_{e^+}+p_{e^-})^2$ are large and comparable,
except for particle masses~\cite{Collins:1989gx}. The original proofs are based
on the study of the analytic structure of Feynman diagrams and on a
power-counting method to find the strength of infrared singularities in massless
perturbation theory.  Factorization is possible because only a limited set of
regions in the space of loop and final state momenta contribute to leading
power, namely the so-called leading regions which are hard, collinear, and soft.
For processes involving fragmentation, the leading regions contain a jet
subdiagram that describes the jet in which the hadron $h$ is
observed~\cite{Collins:1989gx}, see
also~\cite{Mueller:1974yp,Callan:1974rp,Collins:1981uw,Collins:1981uk}.
Accordingly, the fragmentation function that can be constrained by applying
factorization at leading power, corresponds to \eq{defD} only because the sum
over $X$ is dominated by jet-like configurations for the $|X h \rangle$ states.
Therefore, it is interesting to explore whether more can be learned about the
fragmentation process when additional measurements are made on the accompanying
jet.

Here we consider what amounts to the simplest additional measurement, namely
that of the jet invariant mass $m_{Xh}^2=(p_X+p_h)^2$. Rather than using classic
techniques we exploit the powerful computational framework of SCET.

\section{Fragmentation from an Identified Jet in 
$\bar{B}\to Xh \ell\bar\nu$} 
\label{sec:BXhlnu}

Consider the weak transition $b\to u\ell\bar\nu_\ell$ measured with inclusive
decays $\bar B\to X_u\ell\bar\nu_\ell$. The phase space region where $X_u$ is
jet-like plays an important role, because the experimental cuts which remove
$b\to c$ background most often restrict the final state to this region.
Experimentally there is exquisite control over this process, {\it e.g.} in a
large sample of events the neutrino momentum has been reconstructed by
determining the recoil momentum of the $\bar B$, and the spectrum is available
for the jet invariant mass
$m_{X_u}^2$~\cite{Aubert:2004bq,Tackmann:2008qa,:2009tp}.  There has also been
an extensive amount of theoretical work on this process based on the
factorization theorem shown in
\eq{fact4}~\cite{Korchemsky:1999kt,Leibovich:2000ey,Bauer:2001yt,Bauer:2002yu,Bauer:2003pi,Bosch:2004th,Lee:2004ja,Beneke:2004in,Lange:2005yw,Tackmann:2005ub,Ligeti:2008ac}.
From our perspective the nice thing about $\bar B\to X_u\ell\bar\nu_\ell$ is
that it involves only a single jet, and hence provides the simplest possible
framework to extend the factorization analysis involving jet functions to the
fragmentation process we are interested in, where $X_u \to (X h)_u$. Here $h$ is
a light-hadron fragmenting from a $u$-quark, with $m_{h}\ll m_B$.  Without any
loss of generality, we shall refer to $h$ as a pion $\pi$ for the following few
sections, though we will return to the general notation $h$ at the end.

\subsection{Kinematics} \label{sec:kin}

In the $\bar B$ rest frame, the inclusive process $\bar B\to
X_u\ell\bar\nu_\ell$ can be completely described by three variables, often taken
as the hadronic invariant mass $m_{X_u}^2$, the square of the total leptonic
momentum $q^2$ (with $q^\mu=p_\ell^\mu+p_{\bar{\nu}}^\mu$), and the charged
lepton energy $E_\ell$.  In the jet-like region a more convenient set of
variables is $p_{X_u}^+$, $p_{X_u}^-$, and $E_\ell$ where $q^\mu$ is aligned
with the $-\hat z$-axis, and hence the jet-axis is along $+\hat z$ with
$p_{X_u}^\mp = E_{X_u} \pm |\vec p_{X_u}|$.

\begin{figure}[t!]
  \includegraphics*[width=0.7\textwidth]{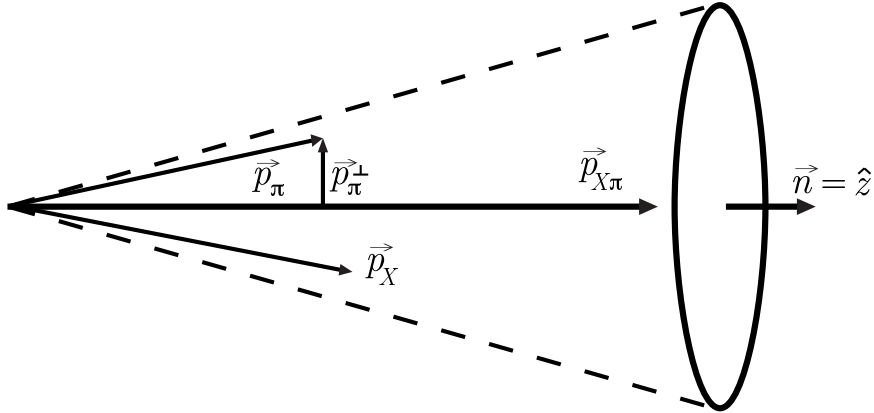}
\vskip-0.3cm
\caption{Kinematic configuration for a pion fragmenting from a jet $X_u\to
  X\pi$.}
\label{cone}
\end{figure}
With an identified hadron in the final state, $\bar B\to X \pi
\ell\bar\nu_\ell$, there are three additional kinematic variables corresponding
{\it e.g.} to the three independent components of $\vec{p}_\pi$. The orientation
of the spatial axes will still be chosen such that $p_{X \pi}^{\perp\mu} = 0$,
as shown in Fig.~\ref{cone}. In this frame the perpendicular component of the
total lepton momentum vanishes, $q^\mu_\perp=0$, and $p_B^\mu=m_B v^\mu$ with
$v^\mu=(n^\mu+\bar{n}^\mu)/2$. The six independent kinematic variables that we
will use to characterize this semi-inclusive process are: $p_{X \pi}^+$, $p_{X
  \pi}^-$, $E_\ell$, $p_{\pi}^-$, $p_{\pi}^+$, $\phi_\ell$, where $\phi_\ell$
denotes the azimuthal angle of the lepton with respect to the $z$-axis and
$p_\pi^\mp= E_\pi \pm p_\pi^z$.  By definition we have $p_{X \pi}^+ \leq p_{X
  \pi}^-$, $E_\pi=(p_\pi^++ p_\pi^-)/2$ and $p_\pi^+=(\vec
p_\pi^{\perp\,2}+m_\pi^2)/p_\pi^-$. In this section we carry out a complete
analysis of the kinematics, determining the phase space limits for the six
kinematic variables without imposing any added restrictions or assumptions.
(Later in Section~\ref{sec:scetsec} we will specialize to the case of a
fragmentation pion collinear in the $\vec{n}$ direction with $p_{\pi}^+ \leq
p_{\pi}^-$.)  Note that $m_{X\pi}^2=p_{X\pi}^- \,p_{X\pi}^+$ so for our process
the measurement of the jet-invariant mass is a measurement of the invariant mass
of all final state hadronic particles.  Lepton masses will be neglected
throughout, but the hadron mass $m_\pi^2$ will be kept for all calculations
involving kinematics.

For three of our six variables we can treat $|X\pi \rangle$ as a combined state $|X_u \rangle$,
and hence $\{p_{X \pi}^+, p_{X \pi}^-, E_\ell\}$ have the same limits as in the
inclusive case, and are given by%
\footnote{Table 2 in Ref.~\cite{Lee:2004ja} lists these limits for $\bar{B} \to
  X_u \ell \bar{\nu}$ for the six possible orders of integration.}
\beq \label{incllimits}
 m_\pi \leq p_{X \pi}^- \leq m_B\;, \;\;\;\;\;\;\;\;\; \frac{m_\pi^2}{p_{X \pi}^-} \leq p_{X \pi}^+ \leq p_{X \pi}^-\;, \;\;\;\;\;\;\;\;\frac{m_B - p_{X \pi}^-}{2} \leq E_\ell \leq \frac{m_B - p_{X \pi}^+}{2}~.
\eeq 

In order to determine the limits for the remaining variables, let us first
consider a frame where $\vec{p}_{X \pi} = 0$, so that it is as if we have $X_u\to
X\pi$ decay in the $X_u$ rest frame. We denote the quantities
evaluated in this frame by a `*'. Since
\beq \label{epibounds}
 E_\pi^*=\frac{m_{X \pi}^2 + m_\pi^2 - p_X^2}{2\, m_{X \pi}}
  \,,
\eeq
the constraint $p_X^2 \geq 0$ implies
\beq
 m_\pi \leq E_\pi^* \leq \frac{m_{X \pi}^2 + m_\pi^2}{2\, m_{X \pi}}~. \label{epilimits}
\eeq
Furthermore, in this $X_u$ rest frame there are no restrictions on the azimuthal
angle $\phi_\pi^*$ of the pion with respect to the $z$-axis, nor of that of the
charged lepton $\phi_\ell^*$, {\it i.e.} $0 \leq \phi_\pi^*, \phi_\ell^* \leq 2
\pi$. The polar angle of the pion is also unconstrained:
\beq
  0 \leq \theta_\pi^* \leq \pi~.
  \eeq

Since ${p_{\pi}^+}^* + {p_\pi^-}^* = 2 E_\pi^*$, from Eq.(\ref{epilimits})
\beq
p_{\pi}^{+\,*}  \leq m_{X \pi} + \frac{m_\pi^2}{m_{X \pi}} - p_\pi^{-\,*}~. \label{eq:ppistarmax}
\eeq
Furthermore, since $|{\vec{p}_\pi}^{\,\perp\,*}|^2 \geq 0$,
\beq
{p_\pi^+}^* \geq \frac{m_\pi^2}{{p_\pi^-}^*}~. \label{eq:ppistarmin}
\eeq
For the maximum $p_{\pi}^{*-}=p_{X\pi}^{*-}=m_{X\pi}$, these limits force
$p_\pi^{*+}=m_\pi^2/m_{X\pi}$ and the pion travels along the $\hat z$ axis.
(Interchanging $p_{\pi}^{*+}\leftrightarrow p_{\pi}^{*-}$ gives the case where
the pion travels along $-\hat z$.) For
$p_{\pi}^{*-}=p_{\pi}^{*+}=(m_{X\pi}+m_\pi^2/m_{X\pi})/2$, we have a pion
traveling purely in the $\perp$-plane with maximal energy. Note that
\eqs{ppistarmax}{ppistarmin} imply other limits such as ${p_\pi^+}^* \leq
|\vec{p}_\pi^{\, \perp\,*}|^2_{ \rm max}\,/ \,p_\pi^{-\,*}$, as well as
${p_\pi^+}^* \leq {p_{X \pi}^+}^* = m_{X \pi}$, and
\beq 
{p_\pi^+}^* = E_\pi^* -
\sqrt{{E_\pi^*}^2 - m_\pi^2}\, \cos{\theta_\pi^*} 
\eeq
for all $\cos{\theta_\pi^*}$. This holds true for both cases
${p_\pi^+}^* \leq {p_\pi^-}^*$ and ${p_\pi^+}^* \geq {p_\pi^-}^*$, which
correspond, respectively, to $0 \leq \cos{\theta_\pi^*} \leq 1$ and $-1 \leq
\cos{\theta_\pi^*} \leq 0$.

Let us now perform a boost along the $z$-axis with velocity $\vec{v}_{X \pi}=
v_{X \pi}\, \hat{e}_z$ to the frame where the $B$-meson decays at rest, which
requires
\beq
 v_{X \pi} = \frac{\sqrt{E_{X \pi}^2 - m_{X \pi}^2}}{E_{X \pi}} =\frac{p_{X \pi}^- - p_{X \pi}^+}{p_{X \pi}^-+p_{X \pi}^+} \label{vxpi}
\eeq
where $0 \leq v_{X \pi} < 1$. Boosting \eqs{ppistarmax}{ppistarmin} yields the
final result for the $p_{\pi}^\pm$ phase space boundaries:
\begin{align}  \label{ppiplbounds}
 & \frac{m_\pi^2}{p_{X \pi}^+} \leq p_\pi^- \leq p_{X \pi}^-~
  \,,
 & \frac{m_\pi^2}{p_\pi^-} & \leq p_\pi^+ \leq p_{X \pi}^+
  \!\left( 1 - \frac{p_\pi^-}{p_{X \pi}^-} \right) 
 + \frac{m_\pi^2}{p_{X \pi}^-}~.
\end{align}
Equivalently, for the opposite order of integration,
\beq  \label{ppireverse}
\frac{m_\pi^2}{p_{X \pi}^-} \leq p_\pi^+ \leq p_{X \pi}^+\, , \;\;\;\;\;\;\; \frac{m_\pi^2}{p_\pi^+} \leq p_\pi^- \leq p_{X \pi}^- \!\left( 1 - \frac{p_\pi^+}{p_{X \pi}^+} \right) + \frac{m_\pi^2}{p_{X \pi}^+}~.
\eeq
Finally, $\phi_\ell^* = \phi_\ell$ since the boost is along the $z$-axis. Hence
\beq
0 \le \phi_\ell \le 2\pi~.
\eeq

\subsection{Differential decay rates} \label{sec:decay}

In this section we derive the fully differential decay rate for $\bar{B}\to
X\pi\ell\bar\nu$ employing only the Lorentz and discrete symmetries of QCD,
without dynamical considerations.  We work in the $B$ rest frame, and it is
convenient to start by using the six independent variables: $q^2$, $E_\ell$,
$E_{\bar {\nu}}$, ${p_\pi}^x$, ${p_\pi}^y$, ${p_\pi}^z$. For the fully
differential decay rate we have
\beq \label{rate}
\frac{d^6 \Gamma}{d q^2\, d E_\ell\, d E_{\bar{\nu}}\, d {p_\pi}^x \, d
  {p_\pi}^y \, d {p_\pi}^z}=\frac{\pi^2}{(2 \pi)^6}\,
  \frac{\cal{A}}{2 E_\pi (2\pi)^3}\: 
  \theta(4 E_\ell E_{\bar{\nu}} - q^2) \,,
\eeq
where $d^3 p_\pi/[ 2 E_\pi(2 \pi)^3]$ is the phase space for the pion,
$E_\pi=\sqrt{\vec{p}_\pi^{\;2}+m_\pi^2}$, and
\beqa \label{A}
{\cal{A}} &\equiv& \sum_{X}\,\sum_{\rm{l.\,s.}}\;\frac{\langle \bar{B} | H_W^\dagger |X \pi \ell \bar{\nu}\rangle\,\langle X \pi \ell \bar{\nu} | H_W | \bar{B}\rangle}{2 m_B}\,(2 \pi)^4 \, \delta^4(p_B - p_{X \pi}  - p_\ell - p_{\bar {\nu}}) \nonumber \\
&=&16 \pi\, G_F^2\, |V_{u b}|^2\, L^{\alpha \beta}\, W_{\alpha \beta} \,.  \eeqa
For the $B$-states we use the relativistic normalization $\langle
\bar{B}(\vec{p}\,) | \bar{B}(\vec{q}\,) \rangle= 2 E_B\, (2 \pi)^3\,
\delta^3(\vec{p}-\vec{q}\,)$.  In Eq.~(\ref{A}) the effective weak Hamiltonian is
\beq
H_W = \frac{4 G_F}{\sqrt{2}}\,V_{u b}\big(\bar{u} \gamma_\mu
P_L b\big)\big(\bar{l} \gamma^\mu P_L
\nu_l\big) \,,
\eeq
where $P_L=(1-\gamma_5)/2$, and factoring the leptonic and hadronic parts of the
matrix element gives the leptonic tensor $L^{\alpha\beta}$, and the hadronic
tensor $W_{\alpha\beta}$.  $L^{\alpha\beta}$ is computed without electroweak
radiative corrections, so $L^{\alpha\beta} = {\rm Tr}[\, \slash\!\!\! p_\ell
\gamma^\alpha P_L\: \slash\!\!\!  p_{\bar\nu} \gamma^\beta P_L ]$. The hadronic tensor
in the $B$ rest-frame in full QCD is 
\beqa  \label{Wmunu} 
W_{\mu \nu} &=& \frac{1}{2 m_B}
\sum_X (2 \pi)^3\, \delta^4(p_B-p_{X \pi}-q) \, \langle \bar{B} |
{J_\mu^u}^\dagger (0) | X \pi\rangle\, \langle X \pi | J_\nu^u
(0) | \bar{B} \rangle \nonumber \\
&=& \frac{1}{4 \pi m_B} \int d^4x\, e^{-i\,q \cdot x}\,\sum_X \,\langle \bar{B}
| {J_\mu^u}^\dagger (x) | X \pi\rangle\, \langle X \pi | J_\nu^u (0) | \bar{B}
\rangle~, 
\eeqa
with the flavor changing weak current $J_\mu^u=\bar{u} \,\gamma_\mu P_L\,b$.  We
have $W_{\alpha\beta}=W_{\alpha\beta}(p_\pi^\mu,v^\mu,q^\mu)$ and we will treat
this tensor to all orders in $\alpha_s$. It can be decomposed using Lorentz
invariance, parity, time reversal, and hermiticity into a sum of scalar
functions, so
\beqa \label{LW}
L^{\alpha \beta} &=& 2\,(p_\ell^\alpha p_{\bar{\nu}}^\beta + p_\ell^\beta
p_{\bar{\nu}}^\alpha - g^{\alpha \beta} \,p_\ell \cdot p_{\bar{\nu}} - i\,
\epsilon^{\alpha \beta \eta \lambda}\, {p_\ell}_\eta\, {p_{\bar{\nu}}}_\lambda)
 \,, \nonumber \\
W_{\alpha \beta}&=& -g_{\alpha \beta}\, W_1 + v_\alpha v_\beta\, W_2
  - i\, \epsilon_{\alpha \beta \mu \nu}\, v^\mu q^\nu \,W_3 
  + q_\alpha q_\beta \,W_4+(v_\alpha q_\beta + v_\beta q_\alpha)\, W_5 
   \nonumber \\ 
  &&  + (v_\alpha {p_\pi}_\beta+v_\beta {p_\pi}_\alpha)\,W_6 
   - i \epsilon_{\alpha \beta \mu \nu}\, p_\pi^\mu q^\nu \,W_7 
   - i\,\epsilon_{\alpha \beta \mu \nu}\, v^\mu p_\pi^\nu \,W_8 
   + {p_\pi}_\alpha {p_\pi}_\beta \,W_9
  \nonumber \\ 
  &&+({p_\pi}_\alpha q_\beta +{p_\pi}_\beta q_\alpha)\, W_{10}~,
\eeqa
with the convention $\epsilon_{0 1 2 3}= 1$.   The scalar functions $W_i$ depend
on the four independent Lorentz invariants $q^2$, $v \cdot q$, $v \cdot p_\pi$
and $p_\pi \cdot q$, or four equivalent variables from our desired set,
\beq
W_i = W_i (p_{X \pi}^+, \,p_{X \pi}^-,\,p_\pi^+,\,p_\pi^-)~. \label{eq:wi}
\eeq
To derive \eq{wi} recall that the leptonic variable $q^\mu$ equals $m_B v^\mu -
p_{X\pi}^\mu $, and can be traded for $p_{X \pi}^\mu$. Also recall that
$m_{X\pi}^2=p_{X\pi}^- \,p_{X\pi}^+$.  Since the $W_i$ do not depend on
$\phi_\ell$ or $E_\ell$ we can (if desired) integrate over these variables
without further information about the functional form of the $W_i$.  In
Eq.~(\ref{LW}) the $W_{i=1-5}$ are analogs of the tensor coefficients that can
appear in the inclusive $\bar{B}\to X_u\ell\bar\nu$ decay, but here they induce
a more differential decay rate because of the identified pion.  The $W_{i=6-10}$
have tensor prefactors involving $p_\pi$ and have no analog in the inclusive
decay.

Contracting leptonic and hadronic tensors we find
\begin{align} \label{lw}
L^{\alpha \beta}\,W_{\alpha \beta} 
 &=  2 q^2\, W_1 + (4 E_\ell E_{\bar{\nu}} - q^2)\,W_2
  +2 q^2\,(E_\ell-E_{\bar{\nu}})\,W_3 
 \nonumber \\
& + (4 E_\ell \,  p_{\bar{\nu}}\cdot p_\pi+4 E_{\bar{\nu}} \, 
p_l \cdot p_\pi -2 E_\pi q^2)\,W_6
 + 2 q^2\,(p_l \cdot p_\pi- p_{\bar{\nu}} \cdot p_\pi)\,W_7  
 \nonumber \\
& +4\,(E_\ell\,p_{\bar{\nu}}\cdot p_\pi- E_{\bar{\nu}}\,p_l\cdot p_\pi)\,W_8 
  +(4\, p_l \cdot p_\pi\, p_{\bar{\nu}} \cdot p_\pi - m_\pi^2 q^2)\,W_9 \,, 
\end{align}
where $W_{4,5,10}$ have dropped out since our leptons are massless. In terms of
this contraction the fully differential decay rate is
\begin{align} \label{rate2}
\frac{d^6 \Gamma}{d q^2\, d E_\ell\, d E_{\bar{\nu}}\, d p_\pi^x\, d p_\pi^y\, d p_\pi^z}
 & =\frac{G_F^2 |V_{ub}|^2}{32 \pi^6}\,
  \frac{L_{\alpha\beta}W^{\alpha\beta}}{2 E_\pi} \,,
\end{align}
where the limits on the kinematic variables are left implicit.

We now want to express Eq.~(\ref{rate2}) in terms of the
coordinates from the previous section:
$\{p_{X\pi}^-,p_{X\pi}^+,E_\ell,p_\pi^-,p_\pi^+,\phi_\ell \}$.  The relations
\begin{align} \label{q2andEn}
  q^2 &= (m_B - p_{X\pi}^-) (m_B - p_{X\pi}^+) \,,
  & E_{\bar\nu} &= m_B - E_\ell - (p_{X\pi}^- + p_{X\pi}^+)/2 \,,
\end{align}
suffice to convert the $W_{1,2,3}$ terms. For the remaining $W_i$ we need
expressions for $p_\ell\cdot p_\pi$, $p_{\bar\nu} \cdot p_\pi$ and $E_\pi$.
Recall that $\vec{p}_{X \pi}=-\vec q = -(\vec{p}_l+\vec{p}_{\bar{\nu}})$ is on
the $+\hat z$-axis, so the leptons are back-to-back in the $\perp$-plane which
is transverse to $\hat z$.  We perform a rotation about the z-axis to bring
$\vec{p}_\pi$ into the $y$-$z$ plane with $p_\pi^y \ge 0$. Then in spherical
coordinates $\vec p_\pi = (p_\pi^x, p_\pi^y, p_\pi^z) = |\vec p_\pi|\, (0,\sin\theta_\pi,
\cos\theta_\pi)$, $\vec p_\ell =E_\ell\, (\sin\theta_\ell \cos\phi_\ell,
\sin\theta_\ell \sin\phi_\ell, \cos\theta_\ell) $, and $\vec p_{\bar\nu} = E_{\bar\nu}\,
(-\sin\theta_{\bar\nu} \cos\phi_\ell, -\sin\theta_{\bar\nu} \sin\phi_\ell,
\cos\theta_\ell)$, so
\begin{align} \label{ppidots}
 p_\ell \cdot  p_{\pi}  &= E_\ell \, E_\pi  -  E_\ell \, |\vec p_\pi|\, 
 (\cos\theta_\ell \cos\theta_\pi + \sin\theta_\pi \sin\theta_\ell \sin\phi_\ell  ) \,,
   \nn\\
  p_{\bar\nu} \cdot  p_{\pi}  &= E_{\bar\nu} \, E_\pi  -  E_{\bar\nu}\, |\vec p_\pi| \,
  (\cos\theta_{\bar\nu} \cos\theta_\pi-\sin\theta_\pi \sin\theta_{\bar\nu} \sin\phi_\ell  ) \,,
   \nn\\ 
 E_\pi &= \frac12 (p_\pi^- + p_\pi^+)\,.
\end{align} 
Using these expressions the two dot products can be written in terms of the
desired variables. First note that
\begin{align} \label{ppicossin}
 |\vec p_\pi| \cos\theta_\pi &= \frac12 (p_\pi^- - p_\pi^+) 
   \,, \\
 |\vec p_\pi |\sin\theta_\pi 
  &= \big( |\vec p_\pi|^2 - |\vec p_\pi|^2\cos^2\theta_\pi \big)^{1/2} 
  =  \big( p_\pi^-  p_\pi^+ -  m_\pi^2  \big)^{1/2}
   \,.\nn 
\end{align}
Furthermore, since $\vec p_\ell \cdot \vec p_{\bar\nu} = E_\ell E_{\bar\nu} - q^2/2$, we have $\vec
p_\ell \cdot \vec q = - E_\ell |\vec q\,| \cos\theta_\ell = \vec p_{\ell}^{\:2}
+ \vec p_\ell \cdot \vec p_{\bar\nu} = E_\ell^2 + E_\ell E_{\bar\nu} -q^2/2$, which, with
$|\vec q\,| = |\vec p_{X\pi}| = [E_{X\pi}^2-m_{X\pi}^2]^{1/2} 
= (p_{X\pi}^- - p_{X\pi}^+)/2$, implies $E_\ell \cos\theta_\ell = {(2E_\ell^2
  \plus 2E_\ell E_{\bar \nu} \minus q^2)}/ {(p_{X\pi}^+ - p_{X\pi}^-)}$.
Hence
\begin{align} \label{EcosEcos}
E_\ell \cos\theta_\ell 
  &= 
  \frac{(m_B\minus p_{X\pi}^-)(m_B\minus p_{X\pi}^+)-
   E_\ell \,(2 m_B\minus p_{X\pi}^- \minus p_{X\pi}^+)} {(p_{X\pi}^- - p_{X\pi}^+)}
  \,,  \\
 E_{\bar\nu} \cos\theta_{\bar{\nu}} 
 &= \frac12(p_{X\pi}^+ - p_{X\pi}^-) - E_\ell \cos\theta_\ell
  \,. \nn
\end{align}
Finally it is useful to note that the equality of the magnitude of the lepton transverse
momenta implies $E_\ell \sin\theta_\ell = E_{\bar \nu}\sin\theta_{\bar\nu}$.

Together the results in Eqs.~(\ref{ppicossin}) and (\ref{EcosEcos}) allow us to
express $L^{\alpha\beta} W_{\alpha\beta}$ in terms of the six variables
$\{E_\ell, p_{X\pi}^\pm, p_\pi^\pm, \phi_\ell \}$.  The only remaining
ingredient needed to transform the decay rate to these variables is the
Jacobian, which is easily derived by noting that
\begin{align}
  \frac{d^3 p_\pi}{2 E_\pi} = \frac14\, dp_\pi^+\, dp_\pi^-\, d\phi_\pi
   =  \frac14 \, dp_\pi^+\, dp_\pi^-\, d\phi_\ell \,.
\end{align}
For the last equality we used the fact that the pion azimuthal angle becomes
equivalent to the lepton azimuthal angle, $d\phi_\pi \to d\phi_\ell$, when we
rotate the pion momentum into the $y$-$z$ plane.  Although it would be
interesting to consider measurements of $\phi_\ell$, for our purposes we will
integrate over $\phi_\ell\in [0, 2\pi]$. Since the $W_{7,8,9}$ prefactors are
linear in $p_\ell\cdot p_\pi$ or $p_{\bar \nu}\cdot p_\pi$ they have
contributions that are either independent of $\phi_\ell$ or linear in
$\sin\phi_\ell$, and the latter terms drop out. In $W_6$ the terms linear in
$\sin\phi_\ell$ do not contribute and a quadratic term averages to $\int_0^{2
  \pi} d\phi_\ell \sin^2(\phi_\ell)= \pi$.  All together this gives
\begin{align} \label{rateEl}
 \frac{d^5 \Gamma}{d p_{X \pi}^+\,d p_{X \pi}^-\,  d p_\pi^- \,d p_\pi^+\, d
   E_\ell}&= \frac{G_F^2\, \vert V_{u b} \vert^2}{128 \pi^5} \Big(\bar{K}_1\,
 W_1 + \bar{K}_2\, W_2 + \bar{K}_3\, W_3 +\bar{K}_6\, W_6+\bar{K}_7\, W_7
 \nonumber \\
 &
\qquad\qquad\qquad +\bar{K}_8\, W_8+\bar{K}_9\, W_9 \Big) ~,
\end{align}
multiplied by $\theta\big[(p_{X \pi}^- + 2 E_\ell-m_B) (m_B-p_{X \pi}^+ - 2
E_\ell)\big]$ which gives the limits for the $E_\ell$ integration, and 
\begin{align}
 \bar{K}_1&= 2\,(p_{X \pi}^- - p_{X \pi}^+) (m_B-p_{X \pi}^-)(m_B-p_{X \pi}^+)
      , \nonumber \\
 \bar{K}_2&= - (p_{X \pi}^- - p_{X \pi}^+)
     (m_B - p_{X \pi}^- - 2 E_\ell)(m_B - p_{X \pi}^+ - 2 E_\ell) 
      , \nonumber \\
 \bar{K}_3&= (p_{X \pi}^- - p_{X \pi}^+)(m_B-p_{X \pi}^-)
  (m_B-p_{X \pi}^+)(4 E_\ell-2 m_B+p_{X \pi}^- + p_{X \pi}^+) 
      , \nonumber \\
 \bar{K}_6&= - 2\,(m_B- 2E_\ell - p_{X \pi}^-)(m_B - 2 E_\ell - p_{X \pi}^+)
    \big[m_B(p_\pi^- - p_\pi^+)+ p_\pi^+\, p_{X \pi}^- - p_\pi^-\, p_{X
      \pi}^+\big] 
  ,\nonumber \\
 \bar{K}_7&= (m_B-p_{X \pi}^-)(m_B - p_{X \pi}^+)
   (4 E_\ell- 2 m_B \plus p_{X \pi}^- \plus p_{X \pi}^+)\big[m_B(p_\pi^-- p_\pi^+) 
   \plus p_\pi^+\, p_{X \pi}^- - p_\pi^-\, p_{X \pi}^+\big] 
   ,\nonumber \\
 \bar{K}_8&= (p_\pi^- - p_\pi^+)(m_B-p_{X \pi}^-)
    (m_B - p_{X \pi}^+)(2 m_B - 4 E_\ell - p_{X \pi}^- - p_{X \pi}^+) 
   ,\nonumber \\
 \bar{K}_9&= \frac{1}{p_{X \pi}^- - p_{X \pi}^+} \,\bigg[
 \Big\{ (p_\pi^- - p_\pi^+)(m_B- p_{X \pi}^-)(m_B- p_{X \pi}^+)
  - 2 E_\ell\,[m_B\,(p_\pi^- - p_\pi^+) 
    \nonumber \\ 
  & +\, p_\pi^+ \,p_{X \pi}^- - p_\pi^-\, p_{X \pi}^+ \big] \Big\} \Big\{m_B^2
  \,(p_\pi^+ - p_\pi^-)+2 m_B\,(p_\pi^- \,p_{X \pi}^+ - p_\pi^+\, p_{X \pi}^-)+
  p_\pi^+ \,{p_{X \pi}^-}^2- p_\pi^- \,{p_{X \pi}^+}^2 
   \nonumber \\ 
  & +\, 2 E_\ell \,\big[m_B (p_\pi^- - p_\pi^+) + p_\pi^+ \,p_{X \pi}^- - p_\pi^-
  \,p_{X \pi}^+\big] \Big\} 
  \nonumber \\
  & +2(m_B-p_{X \pi}^-)(m_B - p_{X \pi}^+)
  (m_B - p_{X \pi}^- - 2 E_\ell)(m_B - p_{X \pi}^+ - 2 E_\ell)(p_\pi^+ p_\pi^- -m_\pi^2) 
  \bigg]
  \nonumber \\ 
  &- 2 m_\pi^2 \,(p_{X \pi}^- - p_{X
    \pi}^+)(m_B- p_{X \pi}^-)(m_B- p_{X \pi}^+) 
   ,
\end{align}
where the limits on the hadronic variables are displayed in
Eqs.~(\ref{incllimits}) and (\ref{ppiplbounds}). The $\bar K_i$ are useful for
considering rates where the pion is observed along with a measurement of the
charged lepton energy.
 
Integrating Eq.~(\ref{rateEl}) over the lepton energy $E_\ell$, the $W_{3,7,8}$
terms drop out leaving
\beq \label{rate1}
\frac{d^4 \Gamma}{d p_{X \pi}^+\,d p_{X \pi}^-\,  d p_\pi^- \,d
  p_\pi^+}=\frac{G_F^2\, \vert V_{u b} \vert^2}{128 \pi^5} \Big( K_1\, W_1
+K_2\, W_2 + K_6\, W_6 + K_9\, W_9 \Big)  \,,
\eeq
where
\begin{align}
K_1&= (m_B- p_{X \pi}^-) (m_B - p_{X \pi}^+)( p_{X \pi}^- - p_{X \pi}^+)^2 
  , \\
K_2 &= \frac{1}{12} (p_{X \pi}^- - p_{X \pi}^+)^4
  ,\nonumber \\
K_6 &= \frac{1}{6} \, (p_{X \pi}^- - p_{X \pi}^+)^3\, 
 \big[ m_B(p_\pi^- - p_\pi^+)+ p_\pi^+ p_{X\pi}^- - p_\pi^- p_{X\pi}^+\big]
  ,\nonumber\\
K_9 &= \frac{1}{12} (p_{X \pi}^- - p_{X \pi}^+)^2 \Big\{
  \big[ p_\pi^+ (m_B- p_{X \pi}^-) + p_\pi^-(m_B- p_{X \pi}^+) \big]^2
  - 4 m_\pi^2 (m_B - p_{X \pi}^-)(m_B- p_{X \pi}^+) 
  \Big\}
. \nn
\end{align}
No further integrations can be performed in Eq.~(\ref{rate1}) without first
determining the hadronic structure functions
$W_i(p_{X\pi}^+,p_{X\pi}^-,p_\pi^+,p_\pi^-)$.

\section{Factorization with a pion fragmenting from a jet.}
\label{sec:scetsec}

Using SCET, we derive a leading order factorization theorem for the hadronic
structure functions $W_i$ appearing in the differential decay rates in
section~\ref{sec:decay}.  

We focus on the region of phase space with endpoint jet-like kinematics where
$p_{X \pi}^+ \ll p_{X \pi}^-$, and with an energetic pion produced by
fragmentation with $p_\pi^+ \ll p_\pi^-$.  It is assumed that suitable
phase-space cuts are applied to subtract the $b \to c$ background, which
phenomenologically is responsible for the importance of this kinematic endpoint
region. This issue is explored in a separate publication~\cite{inprep}. With $p_\pi^+\le p_\pi^-$ the boundaries for $p_\pi^+$ and $p_\pi^-$ in
Eq.~(\ref{ppiplbounds}) become:
\begin{align} \label{ppiboundsSCET}
 & \frac{m_\pi^2}{p_{X
    \pi}^+} \leq p_\pi^- \leq p_{X \pi}^- \,,
 & \frac{m_\pi^2}{p_\pi^-} & \leq p_\pi^+ \leq {\rm min} \left\{ p_\pi^-\, ,
  \,p_{X \pi}^+ \!\left( 1 - \frac{p_\pi^-}{p_{X \pi}^-} \right) +
  \frac{m_\pi^2}{p_{X \pi}^-} \right\}  \,, 
\end{align}
or, reversing the order as in Eq.~(\ref{ppireverse}),
\begin{align}  \label{ppireverseSCET}
 & \frac{m_\pi^2}{p_{X \pi}^-} \leq p_\pi^+ \leq p_{X \pi}^+\,,
 & {\rm max} \left\{ \frac{m_\pi^2}{p_\pi^+},\, p_\pi^+\right\} & \leq p_\pi^- 
 \leq  p_{X \pi}^- \!\left( 1 - \frac{p_\pi^+}{p_{X \pi}^+} \right) +
 \frac{m_\pi^2}{p_{X \pi}^+} ~.
\end{align}

For jet-like final hadronic states, the relevant power counting is $E_{X \pi}
\sim m_b$, $m_b^2 \gg m_X^2 \gtrsim m_b\,\Lambda_{\rm QCD}$, and $p_\pi^- \sim
m_b$. If we decompose the momentum of the remainder of the collinear jet after
the emission of the pion as $p_X ^\mu=(p_X ^+, p_X ^-, p_X^\perp)$, then it
scales as $p_X^\mu \sim (\Lambda_{\rm QCD}, m_b, \sqrt{m_b\, \Lambda_{\rm
    QCD}})=m_b(\lambda^2, 1, \lambda) $ where $\lambda \sim \sqrt{\Lambda_{\rm
    QCD}/m_b}$ is the SCET expansion parameter (which can be defined as
$\lambda^2 = m_{X\pi}^2/m_B^2$ for our process).  The total hadronic momentum
$p_{X \pi}^\mu$ is also collinear and scales the same way as $p_X^\mu$.  We will
start by considering the $(X\pi)$ system as a combined collinear jet, to be
factored from the hard dynamics at the scale $m_b$, and the soft dynamics
responsible for the binding of quarks in the $B$ meson. This part of the
computation can be carried out in \SCETa with collinear and ultrasoft (usoft)
degrees of freedom.  The energetic pion fragments from the jet and has a
collinear scaling $p_\pi^\mu\sim (\lqcd^2/m_b, m_b, \lqcd)$ with much smaller
invariant mass $p_\pi^2 \ll m_{X\pi}^2$. The factorization for this second
fragmentation step can be carried out by a \SCETa to \SCETb matching
computation~\cite{Bauer:2002aj}.

We shall now consider Eq.~(\ref{Wmunu}) at leading order in SCET to derive
factorized expressions for the scalar structure functions $W_i$ in the
fragmentation region.  We work in a frame where $q_\perp=0$, which will induce a
vanishing $\perp$-label momentum for the light quark field in the partonic
subprocess and the $X\pi$ system.  Since the $X\pi$ system is collinear, it is
convenient to decompose momenta as $p^\mu=p_l^\mu + p_r^\mu$ where we have label
momenta $p_l^-\sim \lambda^0$, $p_l^\perp\sim \lambda$, and residual momenta
$p_r^\mu\sim \lambda^2$. In SCET the pion phase space integral can be written as
\begin{align} \label{pionPS}
 \int \frac{d^{3}p_\pi}{2 E_\pi} \,W(\vec p_\pi)
   = \int \frac{dp_\pi^- \, d^2 p_\pi^\perp}{2 p_\pi^-}\, W(p_\pi^-,p_\pi^\perp)
   =  \sum_{p_{\pi l}^-} \,\sum_{p_{\pi l}^\perp} \int \frac{dp_{\pi r}^{-} 
 \, d^2 p_{\pi r}^{\perp}}{2 p_{\pi l}^-} \,W(p_{\pi l}^-,p_{\pi l}^\perp)
 \,.
\end{align}
The same holds for the variables $p_X^-$ and $p_X^\perp$.  Thus for all the
$W=W_i$, we can treat $p_{X\pi}^-$, $p_{X\pi}^\perp$, $p_\pi^-$, and
$p_\pi^\perp$ as discrete label momenta. At the end these variables are restored
to continuous variables using Eq.~(\ref{pionPS}) and the analogs for phase space
integrations over the $X\pi$ variables.

Matching the heavy-to-light QCD current onto SCET operators at a scale of order
$m_b$, at leading order one obtains~\cite{Bauer:2000yr}
\beq \label{matchingcurr}
 J_u^\nu(x) = e^{i {\cal P}\cdot x - i m_b v\cdot x}
 \sum_{j=1}^3 \sum_\omega\, C_j(\omega) \,{J_{uj
    }^\nu}^{(0)}(\omega)  \,.
\eeq
Here ${\cal P}^\mu = n^\mu {\bar {\cal P}}/2 + {\cal P}_\perp^\mu$ where
$\bar{\cal P}$ and $ {\cal{P}}_\perp$ are the ${\cal O}(\lambda^0)$ and ${\cal
  O}(\lambda)$ label momentum operators~\cite{Bauer:2001ct}. The leading
order SCET current becomes
\beq
 {J_{uj}^\nu}^{(0)}(\omega)=\bar{\chi}_{n,\omega}\, \Gamma_{j}^{\nu}
 \,{\cal{H}}_v~.
\eeq
In this expression, $\bar{\chi}_{n,\omega} \equiv \big(\bar{\xi}_{n} \,W_n
\big)\, \delta_{\omega,\bar{\cal{P}}^\dagger}$, where $\xi_n$ is the
$n$-collinear light $u$-quark field.  The collinear Wilson line is defined as
\cite{Bauer:2001ct}
\beq
W_n= \sum_{\rm perms} \exp{\left( -\frac{g}{\bar{{\cal P}}} \,
   \bar{n} \cdot A_{n}(x) \right)}
\eeq
with collinear gluons $A_n$. Also ${\cal{H}}_v \equiv
Y^\dagger h_v$, where $h_v$ is the ultrasoft heavy quark effective theory field,
and $Y(x)=P\exp \big(\,i g \int_{-\infty}^0 ds \,n\cdot A_{us}(ns+x)\big)$ is a
Wilson line built out of ultrasoft gauge fields, which results from decoupling
the ultrasoft gluons from the leading-order collinear
Lagrangian~\cite{Bauer:2001yt}.  For the leading-order Dirac structures we use
the basis \cite{Pirjol:2002km,Chay:2002vy}
\begin{align}
  \Gamma_{1}^{\nu} &= P_R\,\gamma^\nu\,,
  &\Gamma_{2}^{\nu} &=P_R\,v^\nu \,,
  &\Gamma_{3}^{\nu} &=P_R\, \frac{n^\nu}{n \cdot v}\,,
\end{align}
where $P_R=(1+\gamma_5)/2$.  Expressions of the one-loop Wilson coefficients
$C_j$ can be found in Ref.~\cite{Bauer:2001yt} and the two-loop coefficients
were obtained recently by several groups in Refs.~\cite{Bonciani:2008wf,Asatrian:2008uk,Beneke:2008ei,Bell:2008ws}. 
 
\begin{figure}[t!]
  \includegraphics*[width=0.65\textwidth]{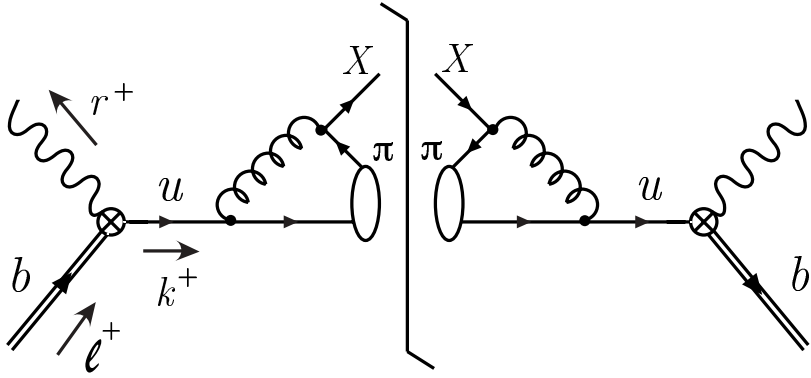}
  \caption{Sketch of the hadronic fragmentation process for $\bar B\to
    X\pi\ell\bar\nu$.} \label{diag}
\end{figure}

For our derivation of the leading-order factorization formula in \SCETa we
follow the steps in Ref.~\cite{Lee:2004ja}, except that we will write out the
dependence on ${\cal P}_\perp$ explicitly.  When the current in Eq.~(\ref{matchingcurr}) is
inserted in Eq.~(\ref{Wmunu}) we have the phase
\begin{align} \label{PSdecomp}
 \int\! {\rm d}^4x\ 
   e^{-i q\cdot x} \,e^{i {\cal P}\cdot x- i m_b v\cdot x} 
  &= \delta_{\bar {\cal P},\bn\cdot p}\ \delta_{{\cal P}_\perp,0} 
   \int\! d^4 x\, e^{- i r\cdot x} \,,
\end{align}
where $\bn\cdot p = m_b - q^-$ and we used the fact that $q_\perp=0$.
Eq.~(\ref{PSdecomp}) leaves discrete $\delta$'s that fix the label momenta, and
a $d^4x$ integration that only involves the residual momentum $r^\mu = \bn^\mu
r^+/2$ with $r^+ = m_b-q^+ \sim \lambda^2$.  Using the normalization convention
in Heavy Quark Effective Theory, $\langle \bar{B}_v(\vec{k}\,') |
\bar{B}_v(\vec{k}\,)\rangle=2 v^0 (2 \pi)^3 \delta^3(\vec{k}-\vec{k}')$, the
leading order expression for Eq.~(\ref{Wmunu}) in SCET becomes
\begin{align} \label{WtensSCET}
 W_{\mu \nu}^{(0)}
 &=\frac{1}{4 \pi} \int d^4x\,e^{-i r\cdot x} 
 \sum_{j,j' =1}^3 \sum_{\omega, \omega' }\, C_{j'}(\omega') \,C_j(\omega)
   \, \delta_{\omega',\bar{n} \cdot p}
   \nonumber \\
 & \times \sum_X 
 \langle \bar{B}_v | \left[ \bar{\cal{H}}_v\, \bar{\Gamma}_{j' \mu}
  \,\chi_{n,\omega',0_\perp}\right]\!(x)| X \pi \rangle 
 \langle X \pi | \left[\bar{\chi}_{n,\omega}\, \Gamma_{j \nu} 
  \,{\cal{H}}_v\right]\!(0) | \bar{B}_v \rangle
 \,,
\end{align}
where ${\chi}_{n,\omega',0_\perp} \equiv
\delta_{\omega,\bar{\cal{P}}}\,\delta_{0,{\cal{P}}_\perp } \big(
W_n^\dagger\, {\xi}_{n}\big)\,$.
Here $\bar{\Gamma}_{j',\mu}=\gamma_0\, \Gamma^{\dagger}_{j'\mu}\gamma_0$ and
$\bn\cdot p = p^-$ is the large momentum of the energetic quark producing the
jet. Grouping ultrasoft and collinear fields by a Fierz transformation, we have
\begin{align} \label{Fierz}
  \left[ \bar{\cal{H}}_v\, \bar{\Gamma}_{j' \mu}
  \,\chi_{n,\omega',0_\perp}\right]\!(x) \left[\bar{\chi}_{n,\omega}\, \Gamma_{j \nu} 
  \,{\cal{H}}_v\right]\!(0)
 &= (-1) \Big[ \bar{\cal{H}}_v(x)  \bar{\Gamma}_{j' \mu}
    \frac{\nslash}{2}\Gamma_{j \nu} {\cal{H}}_v(0)\Big]
    \Big[ \bar{\chi}_{n,\omega}(0) \frac{\bnslash}{4N_c}
    \chi_{n,\omega',0_\perp}(x) \Big] \nn\\
 & + \ldots \,,
\end{align}
where one should keep in mind that the $\langle B_v| \cdots | B_v \rangle$
states will surround the ${\cal H}_v$ bilinear, and the $\langle 0 | \cdots
|X\pi \rangle \langle X\pi|\cdots |0\rangle$ states split the $\chi_{n,\omega}$
field bilinear into two parts. The ellipses in Eq.~(\ref{Fierz}) denote Dirac
and color structures that vanish either because they involve an octet matrix
$T^a$ between the color singlet $|\bar B_v\rangle$ states, or by parity, or
because the only available vector for the $\langle B_v| \cdots | B_v \rangle$
matrix element is $v^\mu$, and $v_\perp^\mu=0$.  The form of the collinear
product of matrix elements is pictured in Fig.~\ref{diag}, and the most general
allowed parameterization is
\begin{align} \label{defGbar}
 \frac{1}{4 N_c}\,{\rm Tr} &\sum_X  \bar{n}\!\!\!/\,\langle 0 | \chi_{n,
   \omega', 0_\perp}(x)| X \pi\rangle  \langle X \pi |
 \bar{\chi}_{n,\omega}(0)|0\rangle = \nonumber \\ &=2\,\delta_{\omega,
 \omega'}\,\delta(x^+) \,\delta^2(x_\perp)\;
  \omega\! \int \frac{d k^+}{2 \pi}\, e^{-i k^+
   x^-/2} \;\bar {\cal G}_u^\pi\Big(k^+ \omega, \,\frac{p_\pi^-}{\omega},
   \,p_\pi^+ p_\pi^-\Big) \,,
\end{align}
where the trace is over color and Dirac indices and Eq.~(\ref{WtensSCET})
implies that $\omega=\bn\cdot p$.  The first $\delta$-function in
Eq.~(\ref{defGbar}) stems from label momentum conservation and the remaining ones
from the fact that the leading collinear Lagrangian contains only the $n \cdot
\partial$ derivative.  The arguments of $\bar {\cal G}_u^\pi$ are constrained by
RPI-III invariance~\cite{Manohar:2002fd}, which requires products of
plus-momenta and minus-momenta, or ratios of minus- (or plus-) momenta. (For our
case RPI-III is equivalent to invariance under boosts along the $\hat z$ jet
axis.)  The arguments of $\bar {\cal G}_u^\pi$ are also constrained by
plus-momentum conservation. The light-cone variable $k^+$ is the plus-momentum
of the up-quark initiating the $X\pi$ production, and at the interaction vertex
is related to the residual (soft) plus-momentum $\ell^+$ of the $b$-quark in the
$B$-meson by $k^+ = \ell^+ - r^+$, as shown in
Fig.~\ref{diag}.\footnote{Strictly speaking the result $k^+ = \ell^+ - r^+$ also
  encodes the presence of Wilson lines in defining these momenta, which ensure
  gauge invariance.}  The large label partonic momentum $\bar{n} \cdot p$ also
is fixed in terms of kinematic variables:
\begin{align}
 \bar{n} \cdot p = m_b - \bar{n}\cdot q=m_b - m_B+\bar{n} \cdot p_{X \pi}=\bar{n} \cdot p_{X \pi}-\bar{\Lambda} + {\cal O}\!\left(\frac{\Lambda_{\rm QCD}^2}{m_b}\right)~.
\end{align}
Since $\bar{\Lambda}={\cal O}(\Lambda _{\rm QCD})$, the ratio $p_\pi^-/p_{X
  \pi}^-$ is identified to leading order with $p_\pi^-/\omega=p_\pi^-/p^- \equiv
z$, the fragmentation variable in \eq{defD}.

Since $p_\pi^+\, p_\pi^- = m_\pi^2 + {\vec{p}_\pi}^{\,\perp\,2}$, $\bar {\cal
  G}_u^\pi$ depends on $|\vec{p}_\pi^{\,\perp}|$, which is non-vanishing with our
choice of coordinates (the pion has $\vec p_\pi^\perp$ and the $\perp$-momentum
of $X$ is $-\vec p_\pi^\perp$).  In general it is the relative $\perp$-momentum
between $\pi$, $X$, and the $B$ that can not be transformed to zero. Later we
will integrate over $|p_\pi^\perp|$ or equivalently $p_\pi^+$, and study
\beq \label{defG}
{\cal G}_u^\pi\!\left(k^+ \omega,\,z, \mu \right) \equiv \omega\! \int d p_\pi^+\; 
{\bar{\cal G}}_u^\pi\!\left(k^+ \omega,\, z,\,p_\pi^+ p_\pi^-, \,\mu\right)~,
\eeq
which occurs in 
\begin{align} \label{defG2}
 \frac{1}{4 N_c}\,{\rm Tr} & \int \!d p_\pi^+ 
  \sum_X \, \bar{n}\!\!\!/\,\langle 0 | \chi_{n,
   \omega', 0_\perp}(x)| X \pi\rangle  \langle X \pi |
 \bar{\chi}_{n,\omega}(0)|0\rangle = \nonumber \\ &=2\,\delta_{\omega,
 \omega'}\,\delta(x^+) \,\delta^2(x_\perp)\;
   \int \frac{d k^+}{2 \pi}\: e^{-i k^+
   x^-/2} \; {\cal G}_u^\pi\Big(k^+ \omega, \,\frac{p_\pi^-}{\omega}\Big) \,.
\end{align}

The \fjet $\bar {\cal G}_u^\pi$ defined in Eq.~(\ref{defGbar}) describes the
properties of a final state that is collimated in the $\vec{n}$-direction and
consists of a up-quark initiated jet from which a pion fragments.  Unlike the
standard unpolarized parton fragmentation function $D_i^\pi(z)$, $\bar {\cal
  G}_u^\pi(s,z,p_\pi^+p_\pi^-)$ carries information about the invariant mass $s$
of the fragmenting jet and the direction of the fragmenting pion through
$p_\pi^+p_\pi^-$. The matrix elements in Eqs.~(\ref{defGbar}) and (\ref{defG2}) are
similar to the collinear matrix element defining the jet function $J(s)$ that
appears in $\bar B\to X_u\ell\bar\nu$ and $\bar{B} \to X_s \gamma$. The jet
function can be written as~\cite{Fleming:2007qr}
\beq \label{J1}
 \frac{1}{4 N_c}\,{\rm Tr} \sum_{X_u} \langle 0 | \bar{n}\!\!\!/\,\chi_{n}(x)|
 X_u \rangle  \langle X_u  | \bar{\chi}_{n,\omega,0_\perp}(0)|0\rangle 
 = \delta(x^+) \,\delta^2(x_\perp)\; \omega\!\! \int d k^+\, e^{-i k^+ x^-/2}\, 
  J_u(\omega k^+) \,.
\eeq
$J_u(s)$ depends on the product $s=k^+\, p_{X_u}^-$, which is analogous to the
first argument in $\bar {\cal G}_u^\pi$.  

The sum in Eq.~(\ref{J1}) extends over states with invariant mass up to
$m_{X_u}^2\sim m_b\,\lqcd$, which are complete in the endpoint region.  Hence
one can write $J$ as the imaginary part (or discontinuity) of a time-ordered
product:
\beq \label{J2}
J_u( k^+\omega)=- \frac{1}{\pi\,\omega}\;{\rm Im} \int d^4 x 
  \; e^{i k \cdot x}\;i\; \big\langle 0\big| \,{\rm T}\
   \!\bar{\chi}_{n,\omega,0_\perp}(0)\,\frac{\bar{n}\!\!\!/}{4 N_c}\, 
  \chi_n(x) \big|0 \big\rangle~,
\eeq
which is perturbatively calculable with
\begin{align}
 J_u(s) = \delta(s)+ {\cal O}(\alpha_s) \,.
\end{align}
On the other hand, $\bar {\cal G}_u^\pi$ involves a pion state, and therefore
contains both perturbative and non-perturbative parts. A simple discontinuity
formula like Eq.(\ref{J2}) does not exist for $\bar{\cal G}_u^\pi$ with the
states $\{ |X\pi\rangle \}$.

Combining Eqs.~(\ref{WtensSCET}) and (\ref{defGbar}) the hadronic tensor at leading
order becomes
\begin{align}  \label{wmn}
 W_{\mu \nu}&=\frac{1}{4 \pi} \int dx^- e^{-i r^+ x^- /2}  \sum_{j,j'=1}^3
 C_{j'}(m_b,p_{X\pi}^-)\, C_j(m_b,p_{X\pi}^-)\,
 {\rm Tr}\left[\frac{P_v}{2} \,\bar{\Gamma}_{j'
     \mu}^{(0)}\, \frac{n\!\!\!/}{2}\,\Gamma_{j \nu}^{(0)} \right] \nonumber \\
 & \times p_{X\pi}^-\!\! \int \frac{d k^+}{2 \pi} e^{-i k^+ x^- /2}
 \ \bar{\cal G}_u^\pi\Big(k^+ p_{X \pi}^-, \,\frac{p_\pi^-}{p_{X \pi}^-}, \,p_\pi^+
   p_\pi^-\Big)\,\langle \bar{B}_v| \bar{h}_v(\tilde{x}) Y(\tilde{x},0)
 h_v(0)|\bar{B}_v \rangle \nn\\
 & \times \left[ 1+ {\cal O}\! \left(\frac{\lqcd^2}{m_{X\pi}^2}\right)\right]
 \,,
\end{align}
where 
\beq
 P_v=\frac{1+v\!\!\!/}{2}\,,\;\;\;\;\;\;\; Y(x,y)=Y(x) Y^\dagger(y)\,,\;\;\;\;\;\; \tilde{x}^\mu=\bar{n} \cdot x \,n^\mu/2~.
\eeq
The matrix element of the bilocal operator in Eq.~(\ref{wmn}) defines the
leading-order shape function~\cite{Neubert:1993um,Bigi:1993ex}
\beq
 f(l^+)=\frac{1}{2}\int \frac{dx^-}{4 \pi}\, e^{-i x^- l^+/ 2} \,\langle \bar{B}_v| \bar{h}_v(\tilde{x}) Y(\tilde{x},0) h_v(0)|\bar{B}_v \rangle=\frac{1}{2} \langle \bar{B}_v |\bar{h}_v\, \delta(l^+-i n\cdot D) \,h_v|\bar{B}_v \rangle~, \label{shapef}
\eeq
with $l^+ = r^++k^+$ for Eq.~(\ref{wmn}). In the limit $m_b \to \infty$, the
support of $f$ is $(- \infty, \bar{\Lambda}]$. The shape function accounts for
non-perturbative soft dynamics in the $B$-meson.  Defining projectors
$P_i^{\mu\nu}$ such that $W_i=W_{\mu \nu}\, P_i^{\mu \nu}$, we obtain the
leading power result
\beq  \label{wi0}
 W_i^{(0)} =\frac{h_i}{\pi} \, p_{X\pi}^- 
 \int_0^{\bar{\Lambda}-r^+}\!\! dk^+\ \bar {\cal G}_u^\pi\Big(k^+ p_{X \pi}^-, 
 \,\frac{p_\pi^-}{p_{X \pi}^-}, \,p_\pi^+ p_\pi^-,\, \mu \Big)\,
  f(k^++r^+,\mu) \,,
\eeq
where we show explicitly the dependence on $\mu$, the renormalization scale.
Here $h_i=h_i(m_b,p_{X\pi}^-,p_{X\pi}^+,\mu)$ where the dependence on
$p_{X\pi}^+$ is entirely from contractions in the tensors, while
that on $m_b,\,p_{X\pi}^-,\,\mu$ comes also from loops.  In terms of the Wilson
coefficients,
\beq \label{hi}
h_i= \sum_{j,j'=1}^3 C_{j'}(m_b,p_{X\pi}^-,\mu)\, C_j(m_b,p_{X\pi}^-,\mu)
 \,{\rm Tr}\left[\frac{P_v}{2} \,\bar{\Gamma}_{j' \mu}^{(0)}\, 
  \frac{n\!\!\!/}{2}\,\Gamma_{j \nu}^{(0)} \right] P^{\mu \nu}_i \,.
\eeq
The projectors $P_i^{\mu \nu}$ relevant for the differential decay rates
in Eqs.~(\ref{rateEl}) and (\ref{rate1}) have the following structure:
\beqa
P_i^{\mu \nu}&=&A_i\,g^{\mu \nu} + B_i\, v^\mu v^\nu + C_i\, q^\mu q^\nu+D_i\,(v^\mu q^\nu+v^\nu q^\mu)+E_i\,p_\pi^\mu \,p_\pi^\nu +F_i\,(v^\mu p_\pi^\nu+v^\nu p_\pi^\mu) + \nonumber \\ && +G_i\,(p_\pi^\mu \,q^\nu+p_\pi^\nu\,q^\mu)+ H_i\,i\,\epsilon^{\mu \nu \alpha \beta}\,v_\alpha \, q_\beta + I_i\,i\, \epsilon^{\mu \nu \alpha \beta} p^\pi_\alpha\, q_\beta + L_i \,i\, \epsilon^{\mu \nu \alpha \beta} v_\alpha p^\pi_\beta~,
\eeqa
where the coefficients $A_i \dots L_i$ are functions of $p_{X \pi}^+$, $p_{X
  \pi}^-$, $p_\pi^+$, $p_\pi^-$ and $m_B$ that are straightforward to determine
by inverting the result for $W_{\alpha\beta}$ in Eq.~(\ref{LW}).

In terms of hadronic variables, Eq.~(\ref{wi0}) becomes
\beqa \label{convol}
 W_i^{(0)}&=&\frac{ h_i}{\pi}\, p_{X\pi}^-\!\!
  \int_0^{p_{X \pi}^+} dk^+ \, \bar{\cal G}_u^\pi\Big(k^+p_{X \pi}^-,\,
   \frac{p_\pi^-}{p_{X \pi}^-},\,p_\pi^+ p_\pi^-,\,\mu\Big)\, 
  f(k^++\bar{\Lambda}-p_{X \pi}^+,\,\mu)
 \nonumber \\
 &=& \frac{h_i}{\pi}\, p_{X\pi}^-\!\! \int_0^{p_{X \pi}^+} dk^+ \, 
  \bar{\cal G}_u^\pi\Big(k^+p_{X \pi}^-,\, \frac{p_\pi^-}{p_{X \pi}^-},\,p_\pi^+
    p_\pi^-, \,\mu\Big)\, S(p_{X \pi}^+ - k^+, \, \mu)
   \nonumber \\
 &=& \frac{h_i}{\pi}\, p_{X\pi}^-\!\! \int_0^{p_{X \pi}^+} dk'^+ \,
  \bar {\cal G}_u^\pi\!\left(p_{X \pi}^-(p_{X \pi}^+ - k'^+),\, 
  \frac{p_\pi^-}{p_{X \pi}^-},\,p_\pi^+ p_\pi^-, \,\mu\right)\, S(k'^+, \, \mu) 
\eeqa
where $S(p) \equiv f(\bar{\Lambda}-p)$ has support for $p\ge 0$. The convolution
variable $k'^+ \equiv p_{X \pi}^+-k^+$ represents the plus-momentum of the
light-degrees of freedom (soft gluons, quarks, and antiquarks) in the $B$-meson,
and $p_{X \pi}^-(p_{X \pi}^+ - k'^+)$ is the invariant mass of collinear
particles in the $u$-quark jet including the fragmentation pion.

Evaluating the traces in Eq.~(\ref{hi}), we derive from Eq.~(\ref{rateEl}) the
following factorization formula for the endpoint fivefold differential decay
rate:
\begin{align} \label{factEl}
\frac{d^5 \Gamma}{d p_{X \pi}^+\, d p_{X \pi}^-\,d p_\pi^- \,d p_\pi^+\,d
  E_\ell}
  &= 3\Gamma_0 \, \bar H(m_B, p_{X \pi}^-, p_{X \pi}^+, E_\ell,\mu) \, p_{X\pi}^-
 \nonumber \\
& \times \int_0^{p_{X \pi}^+} d k'^+\, \bar {\cal G}_u^\pi\Big(p_{X \pi}^-(p_{X
    \pi}^+-k'^+), \frac{p_\pi^-}{p_{X \pi}^-}, p_\pi^+ p_\pi^-,
  \mu\Big)\,S(k'^+,\mu) \,,
\end{align}
with $\Gamma_0\equiv G_F^2\, \vert V_{u b }\vert^2/(1536\, \pi^{6})$ and
\begin{align} \label{HEl}
 & \bar H(m_B, p_{X \pi}^-, p_{X \pi}^+, E_\ell,\mu) 
  = 4\, (m_B - p_{X \pi}^- - 2 E_\ell)\, \Big \{ (m_B-p_{X \pi}^+)(2
  E_\ell-2 m_B +p_{X \pi}^- +p_{X \pi}^+)\, C_1^2
  \nonumber \\ 
  &\qquad
   + (2 E_\ell-m_B +p_{X \pi}^+)\,\Big[ (m_B - p_{X \pi}^+) C_1\, C_2 +
  (p_{X \pi}^- - p_{X \pi}^+) \frac{C_2^2}{4}
  \nonumber \\
  &\qquad
   + 2\, \frac{(m_B-p_{X \pi}^+)^2}{p_{X \pi}^- - p_{X \pi}^+}\, C_1\, C_3
    + (m_B - p_{X \pi}^+)\, C_2\, C_3 + \frac{(m_B-p_{X \pi}^+)^2} 
   {p_{X \pi}^- - p_{X \pi}^+} \, C_3^2  \Big] \Big\} \,.
\end{align}
The $p_{X\pi}^+$- and $E_\ell$-dependence in this expression comes solely from
contraction of the leptonic and hadronic tensors. The renormalized Wilson
coefficients which encode hard loop corrections are functions
$C_i=C_i(m_b,p_{X\pi}^-,\mu)$.

For the decay rate in Eq.~(\ref{rate1}) which integrates over the lepton energy
we obtain: 
\beqa \label{fact1} \frac{d^4 \Gamma}{d p_{X \pi}^+\, d p_{X
    \pi}^-\,d p_\pi^- \,d p_\pi^+} &=&  \Gamma_0\: H(m_B, p_{X \pi}^-,
p_{X \pi}^+,\mu) \, p_{X\pi}^-
\\
&& \times \int_0^{p_{X \pi}^+} d k'^+\, \bar {\cal G}_u^\pi  \Big(p_{X \pi}^-(p_{X
    \pi}^+-k'^+), \frac{p_\pi^-} {p_{X \pi}^-}, p_\pi^+ p_\pi^-,
  \mu\Big)\,S(k'^+,\mu) \,, \nonumber 
\eeqa
where 
\beqa 
H(m_B,p_{X \pi}^-, p_{X \pi}^+,\mu)&=&(p_{X \pi}^- - p_{X \pi}^+)^2
\Big[(m_B-p_{X \pi}^+) \, (3 m_B -2 p_{X \pi}^- - p_{X \pi}^+)\, C_1^2 \nonumber
\\ && +  (m_B-p_{X \pi}^+) (p_{X \pi}^- - p_{X
  \pi}^+) \, C_1\, C_2 
  +(p_{X \pi}^- -p_{X \pi}^+)^2 \, \frac{C_2^2}{4} \nonumber \\ && +
2\, (m_B-p_{X \pi}^+)^2\,C_1\, C_3 + (m_B-p_{X \pi}^+)(p_{X \pi}^- -
p_{X \pi}^+) \, C_2\, C_3\nonumber \\ && + (m_B-p_{X \pi}^+)^2\, C_3^2 \Big]~.  
\eeqa
The functions $\bar H$ and $H$ encode contributions from hard scales, and from
the kinematic contraction of tensor coefficients.  In the phase space region
where $p_{X \pi}^+ \sim \Lambda_{\rm QCD}$, at leading order in the SCET power
counting,
\beq
 H(m_B, p_{X\pi}^-, p_{X\pi}^+,\mu) = H(m_b, p^-,0,\mu) \,.
\eeq
The same considerations apply to the function $H(m_B, p_{X \pi}^-, p_{X \pi}^+,
E_\ell,\mu)$ in Eq.(\ref{factEl}).  Most often it is useful to treat the
$p_{X\pi}^+$ dependence from the tensor contractions exactly, without expanding
$p_{X\pi}^+ \ll p_{X\pi}^-$, since at lowest order in the perturbative
corrections this allows~\cite{Tackmann:2005ub} the endpoint jet-like factorization
theorem to agree with results derived in the more inclusive situation where
$m_{X_u}^2 \sim m_b^2$.

For the purpose of comparison with phenomenology, it is appropriate to derive
the expression of the decay rate which is doubly differential in the jet
invariant mass and in the fraction $z$ of large momentum components. Let us first
integrate Eq.~(\ref{fact1}) over $p_\pi^+$.  At leading order we can set $m_\pi
=0$, since $m_\pi^2 ={\cal O}(\lambda^4)$.  Therefore, in the chiral limit, $
p_\pi^+ \geq 0$ from Eq.~(\ref{ppiboundsSCET}).  Since the pion fragments from the
jet, the maximum value of $p_\pi^+$ is $p_{X \pi}^+-k'^+$. Hence we can write
\beq \label{rate3}
\frac{d ^3 \Gamma}{d p_{X \pi}^+\, d p_{X \pi}^- \,d p_\pi^-} = \Gamma_0\,
H(m_B, p_{X \pi}^-, p_{X \pi}^+,\mu )\, 
 \int_0^{p_{X \pi}^+} dk^+ \,
 {\cal G}_u^\pi\big(k^+p_{X \pi}^-,\,z, \mu \big)\,
 S(p_{X \pi}^+ - k^+,\mu) 
\eeq
where ${\cal G}_u^\pi$ is defined in Eq.~(\ref{defG}). By integrating further: 
\beq \label{double}
\frac{d^2 \Gamma}{d m_{X \pi}^2\,d z} = \int_{m_{X \pi}^2/m_B}^{m_{X \pi}} d
p_{X \pi}^+\;\frac{m_{X \pi}^2}{(p_{X \pi}^+)^2}\; \frac{d^3 \Gamma}{d p_{X
    \pi}^+\,d p_{X \pi}^-\,d p_\pi^-} 
\bigg|_{p_\pi^-=z p_{X\pi}^-}  \bigg|_{p_{X\pi}^-=m_{X\pi}^2/p_{X\pi}^+}
 \,,
\eeq
where we indicate the two changes of variable explicitly. 
The integration boundaries are derived from Eq.~(\ref{incllimits}).

In Sec.\ref{pheno} we will illustrate how to extract from this doubly
differential decay rate information about the standard parton fragmentation
function $D^\pi_u$.

\section{Properties of ${{\cal G}}$} \label{sec:relations}

\subsection{Relations with the inclusive jet function, $J(s,\mu)$}

If we sum over all possible hadrons $h$ in the $X_u\to X h$ fragmentation
process, then the \fjet can be related to the inclusive jet function $J_u(s,\mu)$
which is completely calculable in QCD perturbation theory. 
Consider the equality
\beq
\sum_{h \in {\cal H}_u} \int d p_h^- \int d p_h^+\, \frac{d^4 \Gamma}{d p_{X
    h}^+\,  d p_{X h}^-\,d p_h^- \,d p_h^+} 
  = \frac {d^2 \Gamma}{d p_{X_u}^+\, d p_{X_u}^-}~,
\eeq
where the sum with $h\in {\cal H}_u$ is over all final states with an identified
$h$ hadron fragmenting from the $u$-quark jet. The differential decay rate on
the right-hand side involves the hadronic light-cone variables in the process
$\bar{B} \to X_u \ell \bar{\nu}$. The sum takes $p_{Xh}^+ \to p_{X_u}^+$ and
comparing our Eq.~(\ref{rate3}) with the leading-order factorization theorem for inclusive $\bar{B}\to
X_u\ell\bar\nu$ (see {\it e.g.}~\cite{Ligeti:2008ac}) 
\begin{align}
 \frac {d^2 \Gamma}{d p_{X_u}^+\, d p_{X_u}^-} 
  = 16\pi^3\, \Gamma_0\: H(m_B,p_{X_u}^-,p_{X_u}^+)\, p_{X_u}^- \int_0^{p_{X_u}^+} dk^{\prime\,+}
   J_u\big(p_{X_u}^- (p_{X_u}^+-k^{\prime\,+}),\mu\big) S(k^{\prime\,+},\mu) \,,
\end{align}
one obtains~\cite{nextfrag}:
\beq \label{GtoJ}
 \sum_{h \in {\cal H}_u} \int \!dz \: z\: {\cal G}_j^h\left( k^+ p_{Xh}^-, z, \mu\right) =
 2 \,(2 \pi)^3\,J_j(k^+ p_{X_u}^-,\mu)~, 
\eeq
for $j=u$, where $J_u$ is the leading-order quark jet function. The factor $z$
under the integral is explained in Ref.~\cite{nextfrag}, and is necessary to
provide the correct symmetry factor for states with identical particles. As the
notation indicates, Eq.~(\ref{GtoJ}) holds for other partons $j=\{g,d,\bar
u,\ldots\}$ as well.  This relation between the fragmenting jet function ${\cal
  G}_j^h$ and the jet function is not surprising since the set of states $\{| X
h \rangle_{h \in {\cal H}_u}\}$ is complete.  The factor $2\,(2 \pi)^3$ is
related to how we normalized ${\cal G}_q^h$ and incorporated the phase space for
$h$.

\subsection{Relations with
 the standard fragmentation function $D_q^h(x,\mu)$} \label{sec:pheno}

In the SCET notation, \eq{defD} can be written in terms of the collinear $q$-quark field
\beq
D_q^h \Big(\frac{p_h^-}{\omega}, \mu \Big)=  \pi \omega\! \int d p_h^+
\,\frac{1}{4 N_c}\,{\rm Tr}\,\sum_X\,  \bar{n}\!\!\!/ \,\langle 0|
[\delta_{\omega,\bar{\cal P}}\, \delta_{0,{\cal{P}}_\perp}
 \,\chi_{n}(0)] | X h \rangle \langle X h | \bar{\chi}_{n}(0) |0 \rangle~, \label{Dscet}
\eeq
since $|p_h^\perp|\,d |p_h^\perp|= (p_h^-/2)\,dp_h^+$ at a fixed value of
$p_h^-$. Here $\mu$ is the $\overline {\rm MS}$ renormalization scale.
According to Eq.~(\ref{defG2}), the integral of ${\cal G}_q^h$ over its first
argument can be written as 
\beqa 
\int & &\!\!\!\!\! \frac{d k^+}{2 \pi}\; e^{-i k^+ x^-/2}\; {\cal G}_q^h
\Big(k^+ \omega,\, \frac{p_h^-}{\omega}, \mu \Big) = \nonumber \\ 
&& \!\!\!\!\!\!\!\!\!\!\!\!\!\!= \frac{1}{2} \int d p_h^+ \!\int d x^+ \!\!\int
d^2 x_\perp\, \frac{1}{4 N_c}\,{\rm Tr}\,\sum_X\, \bar{n}\!\!\!/ \,\langle 0|
[\delta_{\omega,\bar{\cal P}}\,\delta_{0,{\cal{P}_\perp}} \chi_{n}
(x)]| X h \rangle \langle X h | \bar{\chi}_{n}(0) |0 \rangle \,.  
\eeqa
If we perform an operator product expansion on the right-hand-side of this
equation we match onto a low energy matrix element that gives the fragmentation function in Eq.~(\ref{Dscet}). Thus, ${\cal G}_j^h$ is given by the
convolution of a perturbatively calculable ${\cal J}_{ij}$ and the standard parton
fragmentation function. The result includes mixing between parton types:
\beq  \label{Gfact}
{\cal G}_i^h\!\left(s,\, z,\mu \right) = \sum_j \int_z^1 \frac{d
  x}{x}\, {\cal J}_{ij}\Big(s,\, \frac{z}{x},\mu \Big) \,
D_j^h(x,\mu) 
 \ \left[1+{\cal O}\Big( \frac{\Lambda_{\rm QCD}^2}{s} \Big)\right]\,, 
\eeq
where $i,j=\{u,d,g,\bar u,\ldots\}$.  In Ref.~\cite{Stewart:2009yx} the concept
of a quark ``beam function'' is discussed. It turns out that a quark beam function is
the analog of ${\cal G}_q^h$, but with parton distributions in place of
fragmentation functions (and an incoming proton in place of an outgoing pion).
The derivation of the factorization theorem in Eq.~(\ref{Gfact}) can be carried
out in a manner analogous to the matching of the gluon beam function onto a
gluon parton distribution, as derived in Ref.~\cite{Fleming:2006cd}.  For the
factorization theorem for the \fjet in Eq.~(\ref{Gfact}), the Wilson coefficient
${\cal J}_{ij}$ describes the formation of a final state jet with invariant mass
$s$ within which the nonperturbative, long-distance fragmentation process takes
place.

At tree level Eq.~(\ref{Gfact}) is easily verified. Using a free $q$-quark of
momentum $p$ in place of $h$ in the final state in Eq.~(\ref{Gfact}) (and
denoting the label parts of $p^\mu$ by $p_\ell^-$, $p_\ell^\perp$ and the
residual parts by $p_r^\mu$, and defining $z=p^-/\omega$), the partonic $\cal
{G}$ is
\begin{align}
 {\cal G}^{\rm tree}_q(k^+ \omega, z)
  &= \int \!\! \frac{dp_r^+}{4} 
 \!\int\!\! dx^- dx^+ d^2x_\perp\, e^{ik^+x^-/2} \frac{1}{4 N_c}
 \, {\rm Tr} \big[ \bnslash\, 
 \delta_{\omega,p_\ell^-}\, \delta_{0,p_\ell^\perp}
 \langle 0 | \xi_n(x) | q(p)\rangle \langle q(p) | \bar\xi_n(0) |0\rangle \big]
 \nn\\
  &= \sum_{p_\ell^\perp} \int\!\! \frac{d^2p_r^\perp}{4\pi p_\ell^-}
   \!\int\!\! dx^- dx^+ d^2x_\perp\, e^{i[(k^+-p_{\ell\perp}^2/p_\ell^-)x^-/2
   -p_r^- x^+/2 - p_r^\perp\cdot x_\perp]}  
  \delta_{\omega,p_\ell^-}\, \delta_{0,p_\ell^\perp} p_\ell^-
\nn\\
 &= \sum_{p_\ell^\perp} \int\!\! \frac{d^2p_r^\perp}{4\pi}
   \, 4 (2\pi)^4 \delta(k^+-p_{\ell\perp}^2/p_\ell^-)\delta(p_r^-)\delta^2(p_r^\perp) 
  \delta_{\omega,p_\ell^-}\, \delta_{0,p_\ell^\perp}
\nn\\
  &= 2 (2\pi)^3 \delta(k^+) \delta(\omega-p^-)
   = 2\, (2\pi)^3 \,\delta(k^+ \omega)\, \delta(1-z)~.
\end{align} 
In the second to last step we recombined the residuals and labels into the
continuous $p^-$, via $\delta_{\omega,p_\ell^-} \delta(p_r^-) = \delta(\omega -
p^-)$.  The quark fragmentation function is ${D}^{\rm tree}_q(z)= \delta(1-z)$.
Since the Wilson coefficients ${\cal J}_{ij}$ are independent of the choice of
states, the tree-level coefficient function can be identified as
\begin{align}
 {\cal J}_{qq}^{\rm tree}(k^+ \omega, z/x,\mu)
  = 2\,(2\pi)^3\,\delta(k^+ \omega)\, \delta(1-z/x) \,,
\end{align}
which satisfies Eq.~(\ref{Gfact}).  The one-loop calculation of ${\cal J}_{ij}$
is presented in Ref.~\cite{nextfrag}.

\section{$D_u^\pi(z)$ from a doubly differential decay rate} \label{pheno}

As a further consequence of our factorization formulae, we explore a strategy to
extract from measurements of suitable differential $B$-decay rates the standard
pion fragmentation function $D_u^\pi(z)$ for values of $z$ that are not too
small, such as $z \gtrsim 0.5$.  Ultimately we anticipate that fragmenting jet
functions will be useful for many other processes (including hadron-hadron
collisions) for which a factorization theorem like Eq.(\ref{fact1}) can be
derived involving ${\cal G}_i^h$.  The phenomenology of $B$-decays is
particularly instructive to this purpose since it allows to concentrate on
single jet production avoiding the kinematical complications of more involved
scattering processes.

For comparison with phenomenology, we are interested in the doubly differential
decay rate in Eq.(\ref{double}). We aim at writing a factorization formula of
the type:
\beq
\frac{d ^2 \Gamma^{\;{\rm cut}}}{d m_{X \pi}^2 \,d z} = {\Gamma}_0\,
\sum_{j=u, \bar{u}, d, g \dots} \int_z^1 \frac{d x}{x}\,\hat{H}_{u j}\left(m_b,m_{X\pi}^2, \frac{z}{x}, \mu \right) \, D_j^\pi(x, \mu) \label{aim}
\eeq
where $\hat{H}_{u j}$ is calculable in perturbation theory and the cut refers to a
suitable interval in $p_{X \pi}^+$ over which we integrate.  We shall argue that
in the ``shape function OPE" regime~\cite{Bauer:2003pi,Bosch:2004th} ($p_{X
  \pi}^- \gg p_{X \pi}^+ \gg \Lambda_{\rm QCD}$) it is possible to write a
factorization formula involving $D_u^\pi$, which does not specify the invariant
mass of the final-state jet.

According to the discussion in Ref.~\cite{Ligeti:2008ac}, the shape function can
be written as a convolution when integrated over a large enough interval
$[0,\Delta]$ such that perturbation theory is applicable at the scale $\Delta$:
\beq
S(\omega) = \int_0^\infty d \omega' \, C_0(\omega-\omega')\,F(\omega')~,
\eeq
where $C_0$ is the $b$-quark matrix element of the shape function operator
calculated in perturbation theory and $F$ is a non-perturbative function that
can be determined by comparison with data. $F$ falls off exponentially for large
$\omega'$ and all its moments exist without a cutoff.

Combining Eqs.~(\ref{rate3}), (\ref{double}) and (\ref{Gfact}), the integration
over $p_{X \pi}^+$ leads to
\beqa
\frac{d ^2 \Gamma^{\;{\rm cut}}}{d m_{X \pi}^2 \,d z}&=&\Gamma_0\,\sum_{j=u, \bar{u}, d, g \dots}\, \int_z^1 \frac{dx }{x}\;D_j^\pi(x, \mu) \int_{m_{X \pi}^2/m_B}^{m_{X \pi}}\,d p_{X \pi}^+\;\frac{m_{X \pi}^2}{(p_{X \pi}^+)^2}\; H\!\left (m_B, \frac{m_{X \pi}^2}{p_{X \pi}^+}, p_{X \pi}^+,\mu \right) \nonumber \\
&& \times  \int_0^{p_{X \pi}^+}d k^+ {\cal J}_{u j}\!\left(k^+ \,\frac{m_{X \pi}^2}{p_{X \pi}^+}, \frac{z}{x}, \mu\right) \int_0^\infty d \omega' \, C_0(p_{X \pi}^+-k^+ -\omega',\mu)\,F(\omega') \label{bigint}
\eeqa
if $p_{X \pi}^{+\,{\rm max}} \gg \Lambda_{\rm QCD}$.  Let us now perform a
Taylor expansion of the perturbative kernel $C_0$ around $\omega'=0$:
\beq
C_0(p_{X \pi}^+-k^+ -\omega')=C_0(p_{X \pi}^+-k^+) - \omega'\,C_0'(p_{X \pi}^+ - k^+) + \dots \label{tay}
\eeq
Since \cite{Ligeti:2008ac}
\beq
\int_0^\infty d \omega'\, F(\omega')=1\;\;\;\;\;\;{\rm and}\;\;\;\;\;\; \int_0^\infty d \omega'\, \omega'^{\,n}\,F(\omega') = {\cal O}(\Lambda_{\rm QCD}^n)~,
\eeq
the $k^+$-convolution integral in Eq.(\ref{bigint}) can be written as
\beq
\int_0^{p_{X \pi}^+} dk^+ \, {\cal J}_{uj}\!\left(k^+ \,\frac{m_{X \pi}^2}{p_{X \pi}^+}, \frac{z}{x}, \mu\right)\,C_0(p_{X \pi}^+-k^+,\mu) + \dots 
\eeq
where the dots indicate terms suppressed by increasing powers of $\Lambda_{\rm
  QCD}/p_{X \pi}^+$ in the phase-space region where the jet becomes less
collimated and increases its invariant mass ($p_{X \pi}^+ \gg \Lambda_{\rm
  QCD}$).  Hence, for $p_{X \pi}^{+\,{\rm min}} \gg \Lambda_{\rm QCD}$ and $p_{X
  \pi}^{+\,{\rm max}} \ll p_{X \pi}^-$, at leading order in the $\Lambda_{\rm
  QCD}/p_{X \pi}^+$-expansion,
\beqa
\frac{d ^2 \Gamma^{\;\rm{cut}}}{d m_{X \pi}^2 \,d z}&=& \Gamma_0\,\sum_{j=u, \bar{u}, d, g \dots}\, \int_z^1 \frac{dx }{x}\;D_j^\pi(x, \mu) \int_{m_{X \pi}^2/m_B}^{m_{X \pi}}\,d p_{X \pi}^+\;\frac{m_{X \pi}^2}{(p_{X \pi}^+)^2}\; H\!\left (m_B, \frac{m_{X \pi}^2}{p_{X \pi}^+}, p_{X \pi}^+,\mu \right) \nonumber \\
&& \times  \int_0^{p_{X \pi}^+}d k^+\, {\cal J}_{u j}\!\left(k^+ \,\frac{m_{X \pi}^2}{p_{X \pi}^+}, \frac{z}{x}, \mu\right)\, C_0(p_{X \pi}^+-k^+,\mu )~.
\eeqa
By identifying
\beqa
\hat{H}_{u j}\!\left(m_B,m_{X \pi}^2,\frac{z}{x},\mu \right)&\Leftrightarrow & \int_{m_{X \pi}^2/m_B}^{m_{X \pi}}\,d p_{X \pi}^+\;\frac{m_{X \pi}^2}{(p_{X \pi}^+)^2}\; H\!\left (m_B, \frac{m_{X \pi}^2}{p_{X \pi}^+}, p_{X \pi}^+,\mu \right) \nonumber \\ && \times \int_0^{p_{X \pi}^+}d k^+ \,{\cal J}_{u j}\!\left(k^+ \,\frac{m_{X \pi}^2}{p_{X \pi}^+}, \frac{z}{x}, \mu\right)\, C_0(p_{X \pi}^+-k^+ ,\mu)~, \label{Htilde}
\eeqa
we see that Eq.~(\ref{aim}) is satisfied at leading order in SCET. Note that to
obtain the complete inclusive $\hat H_{u j}$ there are additional hard corrections
from processes beyond those treated in the jet-like region, so
Eq.~(\ref{Htilde}) does not give the complete expression for $\hat H_{u j}$.
Following the same steps, one can also test the consistency with the factorized
expression
\beq
\frac{d ^3 \Gamma^{\;{\rm cut}}}{d m_{X \pi}^2 \,d z\, dE_\ell} 
 = 3\Gamma_0\,  \int_z^1 \frac{d x}{x}\, {\hat H}_{u j}\!\left(m_b,m_{X \pi}^2, E_\ell, \frac{z}{x}, \mu \right) \, D_j^\pi(x, \mu)~,
\eeq
where $\bar H(m_B, p_{X \pi}^-=m_{X \pi}^2/p_{X \pi}^+, p_{X \pi}^+, E_\ell,
\mu)$ in Eq.~(\ref{HEl}) replaces $H$ in Eq.~(\ref{Htilde}).

\section{Conclusions}  \label{sec:conclusions}
 
Using Soft-Collinear Effective Theory, we have derived leading-order
factorization formulae for differential decay rates in the process $\bar{B} \to
X h \ell \bar{\nu}$ where $h$ is a light, energetic hadron fragmenting from a
measured $u$-quark jet. We obtained results for differential decay rates with
various kinematic variables, for example
\beq 
\frac{d ^3 \Gamma}{d p_{X h}^+\, d p_{X h}^- \,d p_h^-} = \Gamma_0\,
H(m_B, p_{X h}^-, p_{X h}^+,\mu )\, 
 \int_0^{p_{X h}^+} dk^+ \,
 {\cal G}_u^h \big(k^+p_{X \pi}^-,\,z, \mu \big)\,
 S(p_{X \pi}^+ - k^+,\mu)~, 
\eeq
where ${\Gamma}_0$ is a constant prefactor, $H$ encodes contributions from hard
scales, $z = p_h^-/p_{X h}^-$ and $S$ is the leading-order shape function.
${\cal G}_i^h$ is the novel leading-order \fjet: at variance with the standard
parton fragmentation function $D_i^h(z)$, it incorporates information about the
invariant mass of the jet from which the detected hadron fragments.

We have also shown that it is possible to extract $D_i^h(z)$ from a suitable $\bar{B} \to
X h \ell \bar{\nu}$ differential decay rate, for values of $z$ that are not too small, like $z \gtrsim 0.5$.

Moreover, our analysis implies that to obtain a factorization theorem for a
semi-inclusive process where the hadron $h$ fragments from a jet, it is
sufficient to take the factorization theorem for the corresponding inclusive
case and make the replacement
\beq  \label{JtoG}
 J_j(k^+ \omega) \longrightarrow
\frac{1}{2\, (2 \pi)^3}\: {\cal G}_j^h(k^+ \omega, z)~dz \,, 
\eeq 
where $J_j$ is the inclusive jet function for parton $j$ and the additional
phase space variable is $z=p_h^-/p_{Xh}^-$, the momentum fraction of the hadron
relative to the total large momentum of the $Xh$ system.  This replacement
rule is consistent with integration over the phase space for $h$. Applying
Eq.~(\ref{JtoG}) to the factorization theorem given schematically in
Eq.~(\ref{eq:fact4}) we derive the following factorization formula for the
doubly differential decay rate in the process $\bar{B} \to X K \gamma$:
\begin{align} \label{factXK}
&\frac{d^2 \Gamma}{d E_\gamma\, d z}= \frac{\Gamma_{0\, s}\, m_b}{(2 \pi)^3}\, H_s(p_{X K}^+, \mu) \int_0^{p_{X K}^+} d k^+ \,{\cal G}_s^K\!\left(k^+ m_b, z, \mu \right)\,S(p_{X K}^+ - k^+, \mu)  \\
&=\frac{\Gamma_{0\, s}\, m_b}{(2 \pi)^3}\, H_s(p_{X K}^+, \mu)\, \sum_j
\int_0^{p_{X K}^+} d k^+ \!\int_z^1\! \frac{dx}{x}\: {\cal J}_{s j} \!\left(k^+ m_b, \frac{z}{x}, \mu \right) D_j^K(x, \mu)\,S(p_{X K}^+ - k^+, \mu)~, 
\nonumber
\end{align}
with $p_{X K}^+= m_B - 2 E_\gamma$, and the $\Gamma_{0\,s}$ and $H_s$ are
defined in Eq.~(5) and Eq.~(A1) of Ref.~\cite{Ligeti:2008ac}. The soft function
$S$ is the same one as in endpoint $\bar{B}\to X_s\gamma$. The jet Wilson coefficients
${\cal J}_{s j}$ are process independent and calculable in perturbation theory.
At tree level: 
\beq 
{\cal J}_{ss}^{\rm tree}(k^+ \omega, z/x,\mu) =
2\,(2\pi)^3\,\delta(k^+ \omega)\, \delta(1-z/x) ~.  
\eeq

Analogously, for the process $e^+e^- \to \mbox{(dijets)}+h$ we apply
Eq.~(\ref{JtoG}) to the factorization theorem for $e^+e^-\to \mbox{(dijets)}$ in
Eq.~(\ref{eq:fact3}) to obtain the factorized differential cross-section
\begin{align}
  \frac{d^3 \sigma}{d M^2\, d \bar{M}^2\, d z}
  &= \frac{\sigma_0}{2(2 \pi)^3}H_{\rm 2jet}(Q, \mu) 
  \int_{- \infty}^{+ \infty}\!\!\!\! d l^+ d l^-\,
  \Big[ {\cal G}_q^h\!\left(M^2\!-\! Q l^+, z, \mu\right) 
  J_{\bar{n}}\! \left(\bar{M}^2\!-\! Q l^-, \mu \right) + \nn\\
 &\quad\  J_{n}\! \left({M}^2\!-\! Q l^+, \mu \right)
  {\cal G}_{\bar q}^h\!\left(\bar M^2\!-\! Q l^-, z, \mu\right) 
   \Big]
  S_{\rm 2jet}(l^+, l^-, \mu) \nonumber \\
  &= \frac{\sigma_0}{2(2 \pi)^3}H_{\rm 2jet}(Q, \mu) \nn \\
  & \quad \times \sum_j \int_{-
    \infty}^{+ \infty}\! d l^+\, d l^- \int_z^1 \frac{dx}{x}\, \Big[ 
  {\cal J}_{q j}\!\Big(M^2 - Q l^+, \frac{z}{x}, \mu\Big) \,
  J_{\bar{n}}\! \big(\bar{M}^2 -Q l^-, \mu \big)\,
  + \nn\\
 &\quad J_{n}\! \big({M}^2 -Q l^+, \mu \big)
 \,{\cal J}_{\bar q j}\!\Big(\bar M^2 - Q l^-, \frac{z}{x}, \mu\Big) 
  \Big]\:
  D_j^h(x, \mu)\: S_{\rm 2jet}(l^+, l^-, \mu)~,
\end{align}
where $\sigma_0$ is the tree level total cross-section which acts as a
normalization factor, $Q$ is the center-of-mass energy, $M^2$ and $\bar M^2$ are
hemisphere invariant masses for the two hemispheres perpendicular to the dijet
thrust axis. Since here we assume that it is not known whether the hadron $h$
fragmented from the quark or antiquark initiated jet, we have a sum over both
possibilities in the factorization theorem. For the definitions of $\sigma_0$,
$H_{\rm 2jet}$, and $S_{\rm 2jet}$ see Ref.~\cite{Fleming:2007qr} whose
notation we have followed.
 
The factorization formulae derived with our analysis should allow improved
constraints on parton fragmentation functions to light hadrons, by allowing
improved control over the fragmentation environment with the invariant mass
measurement, as well as opening up avenues for fragmentation functions to be
measured in new processes, such as $B$-decays.  We also expect that further
study based on the definition of the \fjet, will contribute to a better
understanding of the relative roles of perturbative partonic short-distance
effects and non-perturbative hadronization in shaping jet properties and
features.
 
\section{Acknowledgments}
This work was supported in part by the Office of Nuclear Physics of the U.S.\ 
Department of Energy under the Contract DE-FG02-94ER40818, and by the Alexander
von Humboldt foundation through a Feodor Lynen Fellowship (M.P.) and a Friedrich
Wilhelm Bessel award (I.S.).  We acknowledge discussions with F.~D'Eramo,
A.~Jain, and W.~Waalewijn.
 
\bibliographystyle{physrev4}
\bibliography{frag}

\end{document}